\documentstyle[12pt,dina4]{article}

\newcommand{\be}{\begin{equation}}
\newcommand{\ee}{\end{equation}}
\newcommand{\bea}{\begin{eqnarray}}
\newcommand{\eea}{\end{eqnarray}}
\newcommand{\beano}{\begin{eqnarrayno}}
\newcommand{\eeano}{\end{eqnarrayno}}
\newcommand{\nnm}{\nonumber}
\newcommand{\const}{\mbox{\rm const}}

\newcommand{\setZ}{{\bf Z}}
\newcommand{\setR}{{\bf R}}
\newcommand{\clos}{\mathop{\rm clos}}
\newcommand{\tr}{\mathop{\rm tr}}
\newcommand{\prll}{\parallel}

\begin{document}

\title{
       \hbox to \hsize{\hfill\vbox{\baselineskip=12pt
                                   \hbox{\small\tt UFTP preprint 357/1994}
                                   \hbox{\small\tt hep-lat/9402012}}}
       Variational description of statistical field theories using Daubechies'
       wavelets
      }

\date{February 16, 1994}

\author{Christoph Best\thanks{email: {\tt cbest@th.physik.uni-frankfurt.de}}
        \ and Andreas Sch\"afer\\[5pt]
        Institut f\"ur Theoretische Physik,\\
        Johann Wolfgang Goethe-Universit\"at,\\
        60054 Frankfurt am Main, Germany}

\maketitle

\begin{abstract}
We investigate the description of statistical field theories using
Daubechies' orthonormal compact wavelets on a lattice. A simple variational
approach is used to extend mean field theory and make predictions for the
fluctuation strengths of wavelet coefficients and thus for the correlation
function. The results are compared to Monte Carlo simulations. We find that
wavelets provide a reasonable description of critical phenomena with only a
small number of variational parameters. This lets us hope for an
implementation of the renormalization group in wavelet space.
\end{abstract}

\section{Introduction}

The success of wavelets
[1-6] in analyzing complex signals has prompted speculation about their
possible applications to field theories and other lattice systems. Like the
Fourier transform, the main feature of the wavelet transform is the
separation of scales. It keeps, however, a (coarse-grained) position-space
lattice that enables one to mix the features of real- and Fourier-space
block-spin transformations, and may thus be a possible candidate for a
renormalization group that could be more efficient both in terms of how
accurate physical processes are depicted and how much numerical work is
necessary to calculate the renormalization group flow. The localization
feature of wavelets is especially appealing when one considers theories that
have a long-range scale (as the confinement scale in Quantum
Chromodynamics). To find out whether a renormalization group description
would work in wavelet space, one must learn about how efficiently a field
theory can be described in terms of wavelets.

In this paper, we study the representation of statistical field theories in
wavelet space on the level of the Gaussian approximation. We use a simple
variational ansatz to calculate fluctuations in wavelet space. These are
related to the correlation function and can be compared to results of Monte
Carlo calculations. Other variational approaches based on the Hamiltonian
formalism \cite{Biro} have been quite successful in Quantum Chromodynamics
(QCD).

In section II, we give a short introduction to the wavelet transform. Our
variational procedure is defined in section III. Its application to the free
field theory and to the $XY$ model is demonstrated in section IV. In section
V, we apply the method to a Landau-Ginzburg field theory.

\section{The wavelet transform}

\subsection{Definition}

A wavelet $\psi \in L^2({\setR})$ is a function whose
binary dilatations and dyadic translations generate a Riesz basis of
$L^2({\setR})$. That is, any function in $L^2(\setR)$ can be expanded into a
wavelet series,
\be
  f(x) = \sum_{n=-\infty}^\infty \sum_{x'\in {\cal L}^n}
  \hat f^{(n)}(x') \, \psi^{(n)}(x')(x) \quad,
\ee
where $\psi^{(n)}(x') \in L^2(\setR)$ denotes the dilatated and
translated wavelet, defined by
\be \label{wvlscale}
  \psi^{(n)}(x')(x) = 2^{-n/2} \psi\left(2^{-n}(x-x')\right) \quad.
\ee
$n\in\setZ$ gives the scale of the wavelet, corresponding to a
dilatation by $2^n$, and $x'\in{\cal L}^n$ the position (translation) on the
sublattice ${\cal L}^n = 2^n\setZ$ of scale $n$.

The coefficients $\hat f^{(n)}(x')$ characterize features of the wavelet
at scale $n$ and position $x'$. In this sense, the wavelet transform offers
a mixed position-frequency representation of the signal.

Compact orthonormal wavelets have been constructed by Daubechies
\cite{DaubP}. They cannot be given in closed form but by a powerful
numerical procedure \cite{NR,Yale1}. The wavelet transform constitutes a
{\em multiresolution analysis} in which the function space $L^2(\setR)$ is
decomposed into a set of nested subspaces
\be
  \cdots \subset V_1 \subset V_0 \subset V_{-1} \subset \cdots
\ee
generated by binary dilatations of a {\em scaling function\/} $\phi(x)$:
\be
  V_n = \clos_{L^2(\setR)} \left\{ \phi^{(n)}(x') : x'\in 2^n\setZ \right\}
\ee
Besides the requirement that their union covers the whole function space and
their intersection is the zero function, they have the important property
that the contraction $f(2x)$ of any function $f\in V_n$ is a member of the
next finer subspace $V_{n+1}$. As it turns out, the pair of scaling function
and wavelet can be chosen such that the spaces $V_n$ are direct sums of
(disjoint) wavelet spaces,
\bea
  W_n &=& \clos_{L^2(\setR)} \left\{ \psi^{(n)}(x') : x'\in 2^n\setZ
          \right\} \quad , \\
  V_n &=& V_{n+1} \oplus W_n \quad.
\eea
Wavelets thus generate a decomposition of the function space into disjoint
subspaces $W_n$, representing different scales. Unlike other methods of
multi-resolution analysis, the wavelet spaces and their scaling function
spaces are derived from a single pair of functions, $\psi(x)$ and
$\phi(x)$, by simple translations and dilatations.

\subsection{Construction}

The construction of wavelets proceeds from the property that the scaling
function $\phi^{(n)}\in V_n$ and the wavelet $\psi^{(n)}\in W_n$ are both
members of the next finer subspace $V_{n-1}$. Thus, they can be expressed as
a combination of the basis function $\{\phi^{(n-1)}(x') : x'\in {\cal
L}^{n-1}\}$ of $V_{n-1}$:
\bea \label{twoscale}
  \phi^{(n)}(k)(x) &=& \sum_l h_l \, \phi^{(n-1)}(k+2^{n-1}l)(x) \nnm\\
  \psi^{(n)}(k)(x) &=& \sum_l g_l \, \phi^{(n-1)}(k+2^{n-1}l)(x)
\eea
This is the {\em two-scale relation\/}. The coefficient
series $\{h_l\}_{l\in\setZ}$ and $\{g_l\}_{l\in\setZ}$ characterize the
wavelet. Compactly supported wavelets have a finite number $L$ of nonvanishing
coefficients. For orthogonality, the coefficient series must be
related by
\be
  g_l = (-1)^l h_{L-l-1} \quad.
\ee
Eq.~(\ref{twoscale}) can be inverted to give ($x'\in {\cal L}^{n-1} =
2^{n-1}\setZ$)
\bea\label{invtwoscale}
  \phi^{(n-1)}(2x')(x) &=& \sum_l \left( h_{2l} \,\phi^{(n)}(2x'-2^n l)(x)
                                + g_{2l} \,\psi^{(n)}(2x'-2^n l)(x) \right)
                                                                      \nnm\\
  \phi^{(n-1)}(2x'+1)(x) &=& \sum_l \big( h_{2l+1} \,\phi^{(n)}(2x'-2^n l)(x)
                                                                      \nnm\\&&
                            \qquad + g_{2l+1} \,\psi^{(n)}(2x'-2^n l)(x) \big)
                                                                 \quad.
\eea

Eq.~(\ref{twoscale}) is the basis of the partial wavelet transform.
If we consider the scalar product of a function $f(x)$ with scaling
functions and wavelets, resp.,
\bea
  \bar f^{(n)}(x') &=& \int \phi^{(n)}(x')(x) \, f(x) \, {\rm d}x \quad,\\
  \hat f^{(n)}(x') &=& \int \psi^{(n)}(x')(x) \, f(x) \, {\rm d}x \quad,
\eea
the coefficients $\bar f^{(n)}(x')$ give a smoothed representation of
$f(x)$, while the wavelet coefficients $\hat f^{(n)}(x')$ contain the
complementary detail information that is lost in the smoothing process.
Using eq.~(\ref{twoscale}), both can be expressed by smoothed coefficients on
the next finer scale,
\bea\label{dwt1}
  \bar f^{(n)}(x') &=& \sum_{l} h_l \bar f^{(n-1)}(x'+2^{n-1}l) \quad,\\
  \hat f^{(n)}(x') &=& \sum_{l} g_l \bar f^{(n-1)}(x'+2^{n-1}l) \quad.
\eea
Repeating the procedure, any wavelet coefficient can be reduced to the
smoothed coefficients at as fine a scale as desired, essentially
corresponding to the resolution of the measurement. Using the inverse
transformation (\ref{invtwoscale}), the basis functions (wavelets and
scaling functions) can be reconstructed as accurately as desired.

Eq.~(\ref{dwt1}) defines an orthogonal transformation between the set
\be \label{wtin}
  \{ \bar f^{(n-1)}(x') : x'\in{\cal L}^{n-1} \}
\ee
of smoothed coefficients at scale $n-1$ and the set of smoothed plus detail
coefficients at scale $n$
\be\label{wtout}
  \{ \hat f^{(n)}(x'), \bar f^{(n)}(x') : x'\in{\cal L}^{n-1} \} \quad.
\ee
If we consider a finite-dimensional vector $(\bar f^{(0)}(x'), x' =
0,\ldots,2^N-1)$ as input (implying periodic continuation where required),
repeated application of this transformation results in the wavelet vector
\be
  \{ \hat f^{(n)}(x'), n=1,\ldots,N, x'\in{\cal L}^n;
     \bar f^{(N)}(0) \}
\ee
As we have only a finite number of elements in the input vector, the partial
wavelet transform must stop when there is only one coefficient left, the
smoothed coefficient of the coarsest scale, giving the average of
$f(x)$. The transformation is orthogonal and thus defines an orthonormal
basis, consisting of discretized wavelets, in the space of vectors. The
scaling relation (\ref{wvlscale}) applies only approximately to these
wavelets. The higher the scale $n$ and the larger the extent of the wavelet,
the less is the discretization felt, and the discrete wavelets approach the
continuum ones.

Wavelets in more than one dimension can be constructed by direct products.
They then carry an additional label $t=1,\ldots,2^D-1$ that specifies their
composition. It is advantageous to put both wavelet and scaling function on
the same footing using an index $t=0,1$,
\be
  \psi^{(1D,n)}_t(x')(x) = \left\{\begin{array}{ll}
                              \phi^{(n)}(x')(x)\quad, & \qquad \mbox{if }t=0 \\
                              \psi^{(n)}(x')(x)\quad, & \qquad \mbox{if }t=1
                           \end{array}\right.
\ee
Then the $N$-dimensional scaling function ($t=0$) and wavelets
($t=1,\ldots,2^D-1$) are given by
\be\label{wvlfac}
  \psi^{(n)}_t(x')(x) = \prod_{k=1}^D
                        \psi^{(1D,n)}_{t_k}(x'_k)(x_k), \qquad
\ee
where
\be
   t = \sum_{k=1}^D t_k 2^{k-1} \quad.
\ee

\section{Method}

\subsection{Wavelet representation}

We study statistical field theories on the lattice in wavelet
representations. The lattice field $a(x)$, $x\in{\cal L}$, is expanded over
a wavelet basis,
\bea
  a(x) &=& \sum_{n=1}^N \sum_t^{n_w} \sum_{x'\in{\cal L}^n}
           \hat a^{(n)}_t(x') \, \psi^{(n)}_t(x')(x)\\
  \hat a^{(n)}_t(x') &=& \sum_{x\in{\cal L}^n}
           \psi^{(n)}_t(x')(x) \, a(x) \quad,
\eea
on a $D$-dimensional lattice ${\cal L}={\cal L}^0$ with $2^{DN}$ sites, $N$
giving the number of scales. The sublattices ${\cal L}^n$ have accordingly
$2^{D(N-n)}$ sites. $t = 0\mbox{ or }1,\ldots,n_w = 2^D-1$ denotes the
cartesian composition of multidimensional wavelets. $t=0$ is included only
at the topmost level $n=N$, where $\phi^{(N)}_0$ is the constant function,
and $\hat a^{(n)}_0$ thus gives the average of $a(x)$ over all sites.

\subsection{Variational principle}

We employ the principle of minimal free energy to obtain an approximate
description of the probability distribution in the partition sum. In a
trial ensemble with probability distribution $P(\alpha;\{a(x)\})$
characterized by a set of variational parameters $\alpha$, we calculate the
entropy $S$
\be
  S = \int \,{\rm d}\{a(x)\} \, P(\alpha,\{a(x)\}) \, \ln P(\alpha,\{a(x)\})
\ee
and the internal energy $U$
\be
  U = \int \,{\rm d}\{a(x)\} \, P(\alpha,\{a(x)\}) \, {\cal H}(\{a(x)\}) \quad.
\ee
By minimizing the free energy
\be
  F = U - \frac{1}{\beta} S
\ee
with respect to the parameters $\alpha$, the best-fit probability
distribution is obtained.

As our goal is to find out how efficiently wavelets describe statistical
field theories, we choose as a trial ensemble the coarsest possibility, a
Gaussian uncorrelated ensemble in wavelet space,
\be \label{uncorr}
  \langle \hat a_{t_1}^{(n_1)}(x_1) \, \hat a_{t_2}^{(n_2)}(x_2) \rangle
  = \delta_{n_1,n_2} \, \delta_{t_1,t_2} \, \delta_{x_1,x_2} \,
    {\cal A}_{t_1}^{(n_1)}
\ee
with scale-dependent fluctuation strengths ${\cal A}_t^{(n)}$. This ensemble
makes calculations sufficiently simple and is, at the same time, able to
describe nonlocal correlations in position space, as we show below. It can
be thought as an extension of the purely local ansatz
\be\label{posansatz}
  \langle a(x) a(y) \rangle = \delta_{x,y} \, {\cal A}
\ee
used in the simplest Gaussian approximation, which corresponds in wavelet
space to constraining all ${\cal A}_t^{(n)}$ to the same value. The
introduction of different fluctuation scales in the wavelet transform hence
enables us to improve the description of nonlocal fluctuations.

The entropy of this ensemble is easily calculated to be
\be\label{entr}
  S = \frac{1}{2} \sum_{n=1}^N \sum_t^{n_w} N_n \, \ln {\cal A}^{(n)}_t \quad
    + \const \quad.
\ee
$N_n = |{\cal L}^n| = 2^{D(N_0-n)}$ is the number of sites on the $n$-th
sublattice.

\subsection{Correlator}

Given the fluctuation strengths ${\cal A}_t^{(n)}$, the correlator can be
calculated by applying the wavelet transform
\be\label{wvlcorr}
  \langle a(x) \, a(y) \rangle =
  \sum_n \sum_t C^{(n)}_t(x,y) \, {\cal A}^{(n)}_{t}
\ee
with the elementary wavelet correlator
\be\label{autcor}
  C^{(n)}_t(x,y) = \sum_{x'\in{\cal L}^n} \psi^{(n)}_t(x')(x) \,
                                          \psi^{(n)}_t(x')(y) \quad.
\ee
These functions are not translation invariant as the translation group has a
complicated representation in wavelet space. Translation invariance of the
correlator (\ref{wvlcorr}) must thus be restored approximately by the ${\cal
A}^{(n)}_t$ in the same way as rotational invariance is restored on a
lattice correlation function.

The scaling relations between wavelets (which hold approximately for
discrete wavelets of sufficiently high scale $n$) imply
\be
  C^{(n)}_t(x,y) \approx C^{(n)}_t(x-y)
  \sim 2^{(n'-n)/2} \, C^{(n')}_t\left(2^{n'-n}(x-y)\right) \quad.
\ee
The elementary wavelet correlators have finite but nonzero extent (as the
wavelets have). Thus, long-range correlations can be obtained by (in wavelet
space uncorrelated) fluctuations at sufficiently large scales. In this way,
wavelets can map a critical position-space theory with long-range
correlations to a less critical wavelet-space theory dominated by
short-range correlations.

\section{Gaussian and $XY$ model}

\subsection{Wavelet fluctuation strengths}

The Gaussian model (free field theory) is given by the Hamiltonian
\be
  {\cal H} = \frac{1}{2} \sum_{x\in{\cal L}} \sum_{\mu=1}^D
             \left( a(x+e_\mu) - a(x) \right)
            +\frac{m^2}{2} \sum_{x\in{\cal L}} \left(a(x)^2\right) \quad.
\ee
It can be solved exactly by a Fourier transform, giving the correlator
\cite{ItzDr}
\be
  \langle a(x) a(x+d) \rangle = \frac{1}{\beta} \, \frac{1}{N_0^2}
    \sum_{k\in{\cal FL}} \frac{
      \cos(k\cdot d)}{
      m^2 + 2\sum_{\mu} ( 2 - 2\cos(k\cdot e_\mu) ) }
\ee
with $k$ on the dual lattice ${\cal FL}$ with lattice spacing $2\pi/2^N$.

The wavelet correlator can be calculated directly by applying the wavelet
transform to this expression:
\be
  \langle \hat a^{(n_1)}_{t_1}(x'_1) \, a^{(n_2)}_{t_2}(x'_2) \rangle
  = \sum_{x_1, x_2 \in{\cal L}} \psi^{(n_1)}_{t_1}(x'_1)(x_1) \,
                                \psi^{(n_2)}_{t_2}(x'_2)(x_2) \,
    \langle a(x_1) \, a(x_2) \rangle \quad.
\ee

On the other hand, fluctuations in wavelet coefficients can be obtained
from the variational principle. To this end, we use (\ref{uncorr}) and
calculate the internal energy in the ensemble. To simplify the
calculation, we rewrite the next-neighbor difference as a scalar product in
lattice field space between the field and the characteristic function of
the link,
\be
  a(x+e_\mu) - a(x) = \langle a, L(x,\mu) \rangle
  = \sum_{y\in{\cal L}} a(y) \, L(x,\mu)(y)
\ee
with the characteristic function
\be\label{linkchi}
  L(x,\mu)(y) = \delta_{y,x+e_\mu} - \delta_{y,x} \quad.
\ee
As the wavelet transform is orthogonal, it conserves the scalar product
\be
  a(x+e_\mu) - a(x) = \langle \hat a, \hat L(x,\mu) \rangle
  = \sum_{n=1}^N \sum_t^{n_w} \sum_{x'\in{\cal L}^n}
    a^{(n)}_t(x') \, \hat L^{(n)}_t(x,\mu)(x')
\ee
where $\hat L^{(n)}_t(x,\mu)(x')$ is the wavelet transform of $L(x,\mu)(y)$.
Then the expectation value of the link Hamiltonian is
\be
  \left\langle \frac{1}{2} \left( a(x+e_\mu) - a(x) \right)^2 \right\rangle_0 =
  \frac{1}{2}
  \sum_{n=1}^N \sum_t^{n_w} |\hat L^{(n)}_t(x,\mu)|^2 \, {\cal A}^{(n)}_t
\ee
with the symbolic notation
\be
  |\hat L^{(n)}_t(x,\mu)|^2 :=
  \sum_{x'\in{\cal L}^n} \left(\hat L^{(n)}_t(x,\mu)(x')\right)^2
\ee
This gives the power spectrum of the link $(x,\mu)$ on scale $n$, in the
sense that the original norm of the function $L(x,\mu)$ is distributed over
different scales, subject to conservation of the norm
\be
  \sum_{n=1}^N \sum_t^{n_w} |\hat L^{(n)}_t(x,\mu)|^2
  = \sum_{y\in{\cal L}} \left(L(x,\mu)(y)\right)^2\quad.
\ee
The notation can be extended to
\be\label{fullat}
  |\hat L^{(n)}_t|^2 := \sum_{x\in{\cal L}} \sum_{\mu=1}^D
                       |\hat L^{(n)}_t(x,\mu)|^2
\ee
which, so to speak, gives the power spectrum of the whole lattice.

Similarly, the site Hamiltonian gives
\be
  \frac{m^2}{2} a(x)^2 =
  \frac{m^2}{2}
  \sum_{n=1}^N \sum_t^{n_w} |\hat S^{(n)}_t(x)|^2 \, {\cal A}^{(n)}_t
\ee
with the characteristic function $S(x)$ of a site:
\be
  S(x)(y) = \delta_{y,x} \quad.
\ee
In this case, there is further analytical simplification as the sum over
$x\in{\cal L}$ can be performed:
\be
  \sum_{x\in{\cal L}} \left(S^{(n)}_t(x)(x')\right)^2
  = \sum_{x\in{\cal L}} \left(\psi^{(n)}_t(x')(x)\right)^2
  = 1
\ee
due to normalization of the wavelets.

The internal energy is therefore
\be
  U = \langle {\cal H} \rangle_0
  = \frac{1}{2} \sum_{n=1}^N \sum_t^{n_w} |\hat L^{(n)}_t|^2 \,
    {\cal A}^{(n)}_t
  + \frac{m^2}{2} \sum_{n=1}^N \sum_t^{n_w} N_n \, {\cal A}^{(n)}_t \quad.
\ee
Both entropy and internal energy are linear in the variational parameters,
so that the variation can be performed analytically, leading to the result
\be\label{gaussA}
  {\cal A}^{(n)}_t = \frac{1}{\beta}\,
  \frac{1}{m^2 + |\hat L^{(n)}_t|^2 / N_n}
\ee
with the notation (\ref{fullat}). The quantity $|\hat L^{(n)}_t|^2/N_n$ has
taken the place of the momentum square $k^2$ in the corresponding
Fourier-space expression.

The wavelet coefficient $n=N, t=0$ gives the average of the function over
all sites. In the case of a characteristic function of a link
(\ref{linkchi}), the average vanishes and thus
\be
  {\cal A}^{(N)}_0 = \frac{1}{\beta \, m^2}
\ee
is divergent for $m^2 \to 0$. The same divergence appears in the Fourier
transform associated with the $k=0$-mode and is caused by the invariance
under adding a constant to a massless field. As it appears only in a single
wavelet coefficient, it can be as easily subtracted in the wavelet case as
with the Fourier transform.

\begin{figure}
\setlength{\unitlength}{0.1bp}
\special{!
/gnudict 40 dict def
gnudict begin
/Color false def
/Solid false def
/gnulinewidth 5.000 def
/vshift -33 def
/dl {10 mul} def
/hpt 31.5 def
/vpt 31.5 def
/M {moveto} bind def
/L {lineto} bind def
/R {rmoveto} bind def
/V {rlineto} bind def
/vpt2 vpt 2 mul def
/hpt2 hpt 2 mul def
/Lshow { currentpoint stroke M
  0 vshift R show } def
/Rshow { currentpoint stroke M
  dup stringwidth pop neg vshift R show } def
/Cshow { currentpoint stroke M
  dup stringwidth pop -2 div vshift R show } def
/DL { Color {setrgbcolor Solid {pop []} if 0 setdash }
 {pop pop pop Solid {pop []} if 0 setdash} ifelse } def
/BL { stroke gnulinewidth 2 mul setlinewidth } def
/AL { stroke gnulinewidth 2 div setlinewidth } def
/PL { stroke gnulinewidth setlinewidth } def
/LTb { BL [] 0 0 0 DL } def
/LTa { AL [1 dl 2 dl] 0 setdash 0 0 0 setrgbcolor } def
/LT0 { PL [] 0 1 0 DL } def
/LT1 { PL [4 dl 2 dl] 0 0 1 DL } def
/LT2 { PL [2 dl 3 dl] 1 0 0 DL } def
/LT3 { PL [1 dl 1.5 dl] 1 0 1 DL } def
/LT4 { PL [5 dl 2 dl 1 dl 2 dl] 0 1 1 DL } def
/LT5 { PL [4 dl 3 dl 1 dl 3 dl] 1 1 0 DL } def
/LT6 { PL [2 dl 2 dl 2 dl 4 dl] 0 0 0 DL } def
/LT7 { PL [2 dl 2 dl 2 dl 2 dl 2 dl 4 dl] 1 0.3 0 DL } def
/LT8 { PL [2 dl 2 dl 2 dl 2 dl 2 dl 2 dl 2 dl 4 dl] 0.5 0.5 0.5 DL } def
/P { stroke [] 0 setdash
  currentlinewidth 2 div sub M
  0 currentlinewidth V stroke } def
/D { stroke [] 0 setdash 2 copy vpt add M
  hpt neg vpt neg V hpt vpt neg V
  hpt vpt V hpt neg vpt V closepath stroke
  P } def
/A { stroke [] 0 setdash vpt sub M 0 vpt2 V
  currentpoint stroke M
  hpt neg vpt neg R hpt2 0 V stroke
  } def
/B { stroke [] 0 setdash 2 copy exch hpt sub exch vpt add M
  0 vpt2 neg V hpt2 0 V 0 vpt2 V
  hpt2 neg 0 V closepath stroke
  P } def
/C { stroke [] 0 setdash exch hpt sub exch vpt add M
  hpt2 vpt2 neg V currentpoint stroke M
  hpt2 neg 0 R hpt2 vpt2 V stroke } def
/T { stroke [] 0 setdash 2 copy vpt 1.12 mul add M
  hpt neg vpt -1.62 mul V
  hpt 2 mul 0 V
  hpt neg vpt 1.62 mul V closepath stroke
  P  } def
/S { 2 copy A C} def
end
}
\begin{picture}(3600,2160)(0,0)
\special{"
gnudict begin
gsave
50 50 translate
0.100 0.100 scale
0 setgray
/Helvetica findfont 100 scalefont setfont
newpath
-500.000000 -500.000000 translate
LTa
600 251 M
0 1758 V
LTb
600 251 M
63 0 V
2754 0 R
-63 0 V
600 780 M
31 0 V
2786 0 R
-31 0 V
600 1090 M
31 0 V
2786 0 R
-31 0 V
600 1309 M
31 0 V
2786 0 R
-31 0 V
600 1480 M
31 0 V
2786 0 R
-31 0 V
600 1619 M
31 0 V
2786 0 R
-31 0 V
600 1737 M
31 0 V
2786 0 R
-31 0 V
600 1839 M
31 0 V
2786 0 R
-31 0 V
600 1929 M
31 0 V
2786 0 R
-31 0 V
600 2009 M
63 0 V
2754 0 R
-63 0 V
600 251 M
0 63 V
0 1695 R
0 -63 V
1040 251 M
0 63 V
0 1695 R
0 -63 V
1480 251 M
0 63 V
0 1695 R
0 -63 V
1920 251 M
0 63 V
0 1695 R
0 -63 V
2361 251 M
0 63 V
0 1695 R
0 -63 V
2801 251 M
0 63 V
0 1695 R
0 -63 V
3241 251 M
0 63 V
0 1695 R
0 -63 V
600 251 M
2817 0 V
0 1758 V
-2817 0 V
600 251 L
LT0
3114 1846 M
180 0 V
600 1867 M
88 -236 V
88 -136 V
88 -93 V
88 -72 V
88 -59 V
88 -51 V
88 -45 V
88 -40 V
88 -37 V
88 -34 V
88 -31 V
88 -29 V
88 -26 V
88 -25 V
88 -23 V
89 -21 V
88 -20 V
88 -18 V
88 -17 V
88 -15 V
88 -14 V
88 -13 V
88 -11 V
88 -10 V
88 -9 V
88 -8 V
88 -7 V
88 -5 V
88 -4 V
88 -3 V
88 -2 V
88 0 V
LT2
3114 1746 M
180 0 V
600 1683 M
88 -368 V
88 -134 V
88 -72 V
88 -51 V
88 -38 V
88 -27 V
88 -25 V
88 -21 V
88 -18 V
88 -15 V
88 -12 V
88 -10 V
88 -10 V
88 -10 V
88 -8 V
89 -7 V
88 -6 V
88 -6 V
88 -5 V
88 -4 V
88 -3 V
88 -3 V
88 -2 V
88 -2 V
88 -1 V
88 -2 V
88 -1 V
88 -1 V
88 0 V
88 0 V
88 0 V
88 0 V
LT3
3114 1646 M
180 0 V
600 1838 M
88 -286 V
88 -124 V
88 -87 V
88 -53 V
88 -54 V
88 -46 V
88 -32 V
88 -28 V
88 -29 V
88 -30 V
88 -27 V
88 -23 V
88 -17 V
88 -14 V
88 -12 V
89 -12 V
88 -12 V
88 -12 V
88 -11 V
88 -11 V
88 -9 V
88 -7 V
88 -6 V
88 -4 V
88 -3 V
88 0 V
88 0 V
88 1 V
88 1 V
88 1 V
88 1 V
88 0 V
LT1
3114 1546 M
180 0 V
600 1870 M
88 -272 V
88 -130 V
88 -101 V
88 -54 V
88 -58 V
88 -64 V
88 -40 V
88 -21 V
88 -28 V
88 -37 V
88 -39 V
88 -32 V
88 -24 V
88 -13 V
88 -6 V
89 -4 V
88 -7 V
88 -9 V
88 -12 V
88 -14 V
88 -13 V
88 -11 V
88 -10 V
88 -7 V
88 -5 V
88 -2 V
88 1 V
88 2 V
88 3 V
88 3 V
88 3 V
88 0 V
stroke
grestore
end
showpage
}
\put(3054,1546){\makebox(0,0)[r]{DAUB20}}
\put(3054,1646){\makebox(0,0)[r]{DAUB8}}
\put(3054,1746){\makebox(0,0)[r]{DAUB4}}
\put(3054,1846){\makebox(0,0)[r]{exact}}
\put(2008,2109){\makebox(0,0){Gaussian correlator, $m=0.032$, $64\times 64$}}
\put(2008,51){\makebox(0,0){$d$}}
\put(100,1130){%
\special{ps: gsave currentpoint currentpoint translate
270 rotate neg exch neg exch translate}%
\makebox(0,0)[b]{\shortstack{$C(d)$}}%
\special{ps: currentpoint grestore moveto}%
}
\put(3241,151){\makebox(0,0){30}}
\put(2801,151){\makebox(0,0){25}}
\put(2361,151){\makebox(0,0){20}}
\put(1920,151){\makebox(0,0){15}}
\put(1480,151){\makebox(0,0){10}}
\put(1040,151){\makebox(0,0){5}}
\put(600,151){\makebox(0,0){0}}
\put(540,2009){\makebox(0,0)[r]{1}}
\put(540,251){\makebox(0,0)[r]{0.1}}
\end{picture}
\caption{Correlator of a free field at $\beta=1$. The solid line is the
exact result, the broken lines the result of the variational procedure with
different wavelet types.}
\label{fig7}
\end{figure}
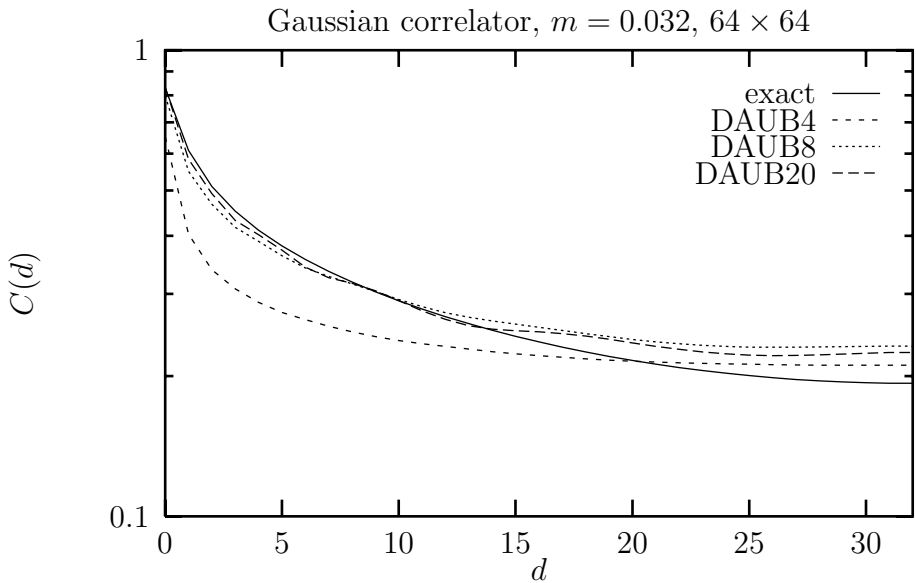

Fig.~\ref{fig7} shows the prediction for the correlator from different
wavelets. The higher the order of the wavelet, the more correct the result.
This is not surprising as higher wavelets approach more and more the form of
free waves that diagonalize the Hamiltonian.

\subsection{Correlator}

To calculate the correlator of the massless theory, we first derive the
scaling properties of the fluctuation strengths ${\cal A}^{(n)}_t$ in
respect to $n$. Using the factorization property (\ref{wvlfac}) of
the wavelets, we find for (\ref{fullat})
\bea
  \hat L^{(n)}_t(x,\mu)(x')
  &=& \psi^{(n)}_t(x+e_\mu-x') - \psi^{(n)}_t(x-x') \nnm\\
  &=& \left( \prod_{k\ne \mu}^D \psi^{(1D,n)}_{t_k}(x_k-x'_k) \right) \nnm\\
  &&\quad\times
    \left( \psi^{(1D,n)}_{t_k}(x_\mu+1-x'_\mu) - \psi^{(1D,n)}_{t_k}(x_\mu)
    \right)
\eea
Summing over $x'$, $x$, and $\mu$ yields
\be
  |\hat L^{(n)}_t|^2 =
  2^{D(N-n)} D \sum_x \left(
    \psi^{(1D,n)}_{t_k}(x_\mu+1) - \psi^{(1D,n)}_{t_k}(x_\mu)
  \right)^2 \quad.
\ee
The one-dimensional finite difference can be approximated by a
derivative as the discrete wavelets approach the continuum wavelets
(\ref{wvlscale}) for large $n$
\bea &&
  \psi^{(1D,n)}_{t_k}(x_\mu+1) - \psi^{(1D,n)}_{t_k}(x_\mu)
  \nnm\\
  &\approx&
  2^{-n/2} \left( \psi(2^{-n}(x+1)) - \psi(2^{-n}x) \right)
  \approx
  2^{-3n/2} \psi^\prime(2^{-n}x) \quad.
\eea
Consequently
\bea &&
  \sum_x \left(
    \psi^{(1D,n)}_{t_k}(x_\mu+1) - \psi^{(1D,n)}_{t_k}(x_\mu)
  \right)^2 \nnm\\
  &\approx& 2^{-3n} \sum_x \left(\psi^\prime(2^{-n}x)\right)^2
  \approx   2^{-2n} \int{\rm d}x\, \left(\psi^\prime(x)\right)^2
\eea
and we deduce the scaling relation
\be \label{Lscal}
  |\hat L^{(n)}_t|^2 \sim 2^{-(D+2)n} \qquad\Longrightarrow\qquad
  {\cal A}^{(n)}_t \sim \frac{4^n}{2\beta} \quad.
\ee
We also note here that, in the massive case $m\ne 0$, (\ref{gaussA})
implies that the ${\cal A}^{(n)}_t$ tend to the constant value $1/\beta m$.
As we shall see, this corresponds to the absence of long-range correlations.

For the wavelet autocorrelation $C^{(n)}_t(x,y)$ function, we capture its
gross features by
\be
  C^{(0)}_t(x,x+d) \approx (1-2d) \, \Theta(1-d) \, \Theta(d) \quad.
\ee
The correlation function of the field theory is then
\be
  \langle a(x) a(x+d) \rangle =
  \sum_{n=1}^N \sum_t^{n_w} {\cal A}^{(n)}_t
  (1 - 2^{1-n} d) \, \Theta(1 - 2^{1-n} d)\, \Theta(d) \quad.
\ee
Let $2^{n_0} < d < 2^{n_0+1}$. Then only terms $n\ge n_0$ contribute in the
sum. If ${\cal A}^{(n)}_t \sim \alpha^n {\cal A}^{(0)}$, the sum can be
performed, sending $N\to\infty$,
\bea
  C(d) &=& {\cal A}^{(0)} \, n_w \, \Theta(d) \,
  \sum_{n=n_0}^\infty \alpha^n \, 2^{-Dn} \, (1-2^{1-n}d) \nnm\\
  &=& {\cal A}^{(0)} \, n_w \, \Theta(d) \,
  \alpha^{n_0} \, 2^{-Dn_0} \,
  \left( \frac{2^D}{2^D-\alpha} - 2^{1-n_0} d
                   \frac{2^{D+1}}{2^{D+1}-\alpha} \right) \quad.
\eea
Putting $2^{n_0} \sim d$ and thus $\alpha^{n_0} \sim d^{\log_2\alpha}$, we
find
\be
  C(d) \sim d^{-(D-2)} \quad.
\ee
This is the expected scaling behavior for a Gaussian theory \cite{ItzDr}. In
two dimensions, it corresponds to a logarithmic correlation function.

\subsection{XY model}

We would like to test our approximation on a nontrivial field theory. The
two-dimensional $XY$ model is among the simplest field theories with a
nontrivial critical behavior.

The Hamiltonian of the $XY$ model is
\be
  {\cal H} = \sum_{x\in{\cal L}} \sum_{\mu=1}^D
             \cos\left( a(x+e_\mu) - a(x) \right) \quad.
\ee
The field $-\pi \le a(x) <\pi$ gives the angle of the spin at $x$, relative
to some global reference direction. It is symmetric under global $O(1)$
transformations and undergoes a Kosterlitz-Thouless phase transition in two
dimensions \cite{ItzDr}.

As in the Gaussian model, we expand the field $a(x)$ in a wavelet basis,
assume Gaussian local fluctuations, and calculate the internal energy as the
expectation value of the Hamiltonian in the ensemble. However, to make a
Gaussian ansatz, we have to extend the field $a(x)$ to the whole real line.
The Gaussian ansatz now does not reflect the periodicity of the Hamiltonian.
This is a valid approximation in the ordered low-temperature phase, but it
fails at high temperatures. This fact is, however, independent of the
wavelet expansion.

To calculate the internal energy, it is useful to note that
\be
  \left\langle \cos( x + x_0 ) \right\rangle
  = \exp\left(-\frac{\langle x^2\rangle}{2} \right) \, \cos x_0
\ee
for a Gaussian variable $x$. With this and the same procedure as above, we
find for the internal energy
\be \label{xyU}
  U = \sum_{x\in{\cal L}} \sum_{\mu=1}^D \left[ 1 -
  \exp\left( -\frac{1}{2} \sum_n^N \sum_t^{n_w} \, |L^{(n)}_t(x,\mu)|^2 \,
             {\cal A}^{(n)}_t \right) \right ]\quad.
\ee
In the limit of small fluctuations, the exponential can be expanded, and
the (massless) Gaussian model is recovered for the low-temperature limit.
Conversely, at high temperatures the exponential vanishes and the internal
energy approaches unity per site, reflecting maximal disorder.

The free energy exhibits in this approximation a local minimum that vanishes
when the temperature increases. The absence of a global minimum is rooted in
the lack of periodicity of the Gaussian probability distribution. While the
internal energy $U$ is bounded by unity at maximal disorder, our entropy
(\ref{entr}) can increase without bounds. The true entropy, however, is
bounded by the available configuration space, $S_{\rm max} = N_0 \ln 2\pi$.
By assuming a Gaussian probability distribution, we have extended the
configuration space to the infinite real line.

We can, however, still interpret this result from the point of view of
ergodicity in a system with a periodic potential: In the low-temperature
phase, the system is localized in a small part of the available
configuration space. When the temperature increases, the localized solution
becomes unstable at some point, but as our trial probability distribution
space does not include periodic distributions, the minimization problem has
no solution.

Numerically, we find the temperature of this phase transition at $\beta_{\rm
crit} \approx 0.69$ in two dimensions and $\beta_{\rm crit}\approx 0.45$ in
three dimensions. This is consistent both with the result for the purely
local ansatz (\ref{posansatz}), which gives $\beta_{\rm crit} = e/2D \approx
0.6769\ldots$ in two dimensions and $0.4530\ldots$ in three dimensions, and
with the prediction from an analysis of vortices in two
dimensions\cite{ItzDr} giving $\beta_{\rm crit} = 2/\pi \approx
0.6366\ldots$.


Wavelets, however, do not only predict the internal energy but also nonlocal
correlations that are not included in the position-space ansatz.
Fig.~\ref{fig5} compares the lowest wavelet fluctuation strengths as
calculated variationally and as measured in a Monte Carlo simulation. This
essentially amounts to comparing correlators, which are not predicted by
(\ref{posansatz}). In the Monte Carlo simulation, we have ignored
periodicity (i.e.~allowed $a(x)$ to be outside the interval $[-\pi,\pi]$)
to model the Gaussian ansatz as closely as possible.

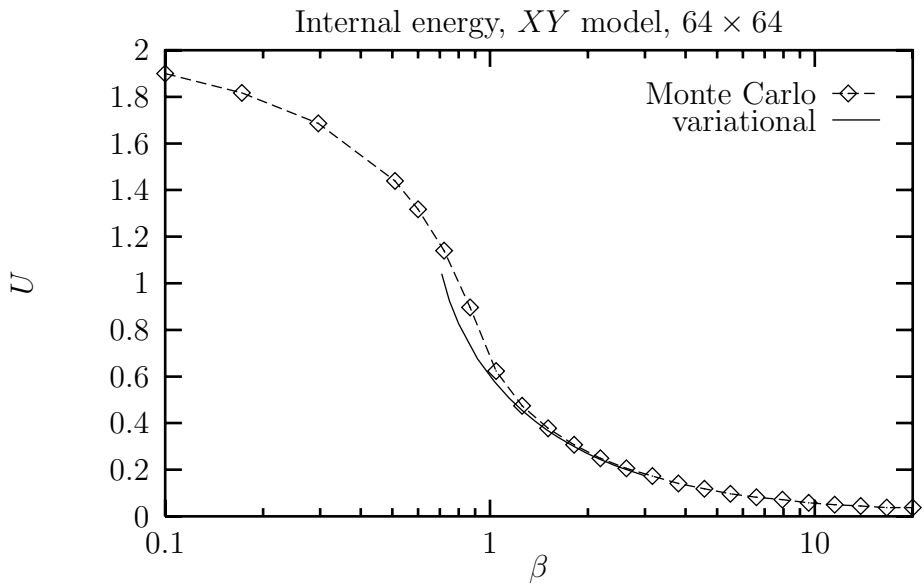
\begin{figure}
\setlength{\unitlength}{0.1bp}
\special{!
/gnudict 40 dict def
gnudict begin
/Color false def
/Solid false def
/gnulinewidth 5.000 def
/vshift -33 def
/dl {10 mul} def
/hpt 31.5 def
/vpt 31.5 def
/M {moveto} bind def
/L {lineto} bind def
/R {rmoveto} bind def
/V {rlineto} bind def
/vpt2 vpt 2 mul def
/hpt2 hpt 2 mul def
/Lshow { currentpoint stroke M
  0 vshift R show } def
/Rshow { currentpoint stroke M
  dup stringwidth pop neg vshift R show } def
/Cshow { currentpoint stroke M
  dup stringwidth pop -2 div vshift R show } def
/DL { Color {setrgbcolor Solid {pop []} if 0 setdash }
 {pop pop pop Solid {pop []} if 0 setdash} ifelse } def
/BL { stroke gnulinewidth 2 mul setlinewidth } def
/AL { stroke gnulinewidth 2 div setlinewidth } def
/PL { stroke gnulinewidth setlinewidth } def
/LTb { BL [] 0 0 0 DL } def
/LTa { AL [1 dl 2 dl] 0 setdash 0 0 0 setrgbcolor } def
/LT0 { PL [] 0 1 0 DL } def
/LT1 { PL [4 dl 2 dl] 0 0 1 DL } def
/LT2 { PL [2 dl 3 dl] 1 0 0 DL } def
/LT3 { PL [1 dl 1.5 dl] 1 0 1 DL } def
/LT4 { PL [5 dl 2 dl 1 dl 2 dl] 0 1 1 DL } def
/LT5 { PL [4 dl 3 dl 1 dl 3 dl] 1 1 0 DL } def
/LT6 { PL [2 dl 2 dl 2 dl 4 dl] 0 0 0 DL } def
/LT7 { PL [2 dl 2 dl 2 dl 2 dl 2 dl 4 dl] 1 0.3 0 DL } def
/LT8 { PL [2 dl 2 dl 2 dl 2 dl 2 dl 2 dl 2 dl 4 dl] 0.5 0.5 0.5 DL } def
/P { stroke [] 0 setdash
  currentlinewidth 2 div sub M
  0 currentlinewidth V stroke } def
/D { stroke [] 0 setdash 2 copy vpt add M
  hpt neg vpt neg V hpt vpt neg V
  hpt vpt V hpt neg vpt V closepath stroke
  P } def
/A { stroke [] 0 setdash vpt sub M 0 vpt2 V
  currentpoint stroke M
  hpt neg vpt neg R hpt2 0 V stroke
  } def
/B { stroke [] 0 setdash 2 copy exch hpt sub exch vpt add M
  0 vpt2 neg V hpt2 0 V 0 vpt2 V
  hpt2 neg 0 V closepath stroke
  P } def
/C { stroke [] 0 setdash exch hpt sub exch vpt add M
  hpt2 vpt2 neg V currentpoint stroke M
  hpt2 neg 0 R hpt2 vpt2 V stroke } def
/T { stroke [] 0 setdash 2 copy vpt 1.12 mul add M
  hpt neg vpt -1.62 mul V
  hpt 2 mul 0 V
  hpt neg vpt 1.62 mul V closepath stroke
  P  } def
/S { 2 copy A C} def
end
}
\begin{picture}(3600,2160)(0,0)
\special{"
gnudict begin
gsave
50 50 translate
0.100 0.100 scale
0 setgray
/Helvetica findfont 100 scalefont setfont
newpath
-500.000000 -500.000000 translate
LTa
600 251 M
2817 0 V
LTb
600 251 M
63 0 V
2754 0 R
-63 0 V
600 427 M
63 0 V
2754 0 R
-63 0 V
600 603 M
63 0 V
2754 0 R
-63 0 V
600 778 M
63 0 V
2754 0 R
-63 0 V
600 954 M
63 0 V
2754 0 R
-63 0 V
600 1130 M
63 0 V
2754 0 R
-63 0 V
600 1306 M
63 0 V
2754 0 R
-63 0 V
600 1482 M
63 0 V
2754 0 R
-63 0 V
600 1657 M
63 0 V
2754 0 R
-63 0 V
600 1833 M
63 0 V
2754 0 R
-63 0 V
600 2009 M
63 0 V
2754 0 R
-63 0 V
600 251 M
0 63 V
0 1695 R
0 -63 V
969 251 M
0 31 V
0 1727 R
0 -31 V
1184 251 M
0 31 V
0 1727 R
0 -31 V
1337 251 M
0 31 V
0 1727 R
0 -31 V
1456 251 M
0 31 V
0 1727 R
0 -31 V
1553 251 M
0 31 V
0 1727 R
0 -31 V
1635 251 M
0 31 V
0 1727 R
0 -31 V
1706 251 M
0 31 V
0 1727 R
0 -31 V
1768 251 M
0 31 V
0 1727 R
0 -31 V
1824 251 M
0 63 V
0 1695 R
0 -63 V
2193 251 M
0 31 V
0 1727 R
0 -31 V
2408 251 M
0 31 V
0 1727 R
0 -31 V
2561 251 M
0 31 V
0 1727 R
0 -31 V
2680 251 M
0 31 V
0 1727 R
0 -31 V
2777 251 M
0 31 V
0 1727 R
0 -31 V
2859 251 M
0 31 V
0 1727 R
0 -31 V
2930 251 M
0 31 V
0 1727 R
0 -31 V
2992 251 M
0 31 V
0 1727 R
0 -31 V
3048 251 M
0 63 V
0 1695 R
0 -63 V
3417 251 M
0 31 V
0 1727 R
0 -31 V
600 251 M
2817 0 V
0 1758 V
-2817 0 V
600 251 L
LT0
LT1
3114 1846 M
180 0 V
600 1921 M
289 -73 V
288 -115 V
289 -217 V
87 -107 V
98 -156 V
98 -214 V
98 -240 V
98 -131 V
98 -85 V
98 -62 V
99 -51 V
98 -38 V
98 -29 V
98 -28 V
98 -20 V
98 -19 V
98 -13 V
98 -9 V
99 -12 V
98 -7 V
98 -5 V
98 -6 V
98 0 V
3174 1846 D
600 1921 D
889 1848 D
1177 1733 D
1466 1516 D
1553 1409 D
1651 1253 D
1749 1039 D
1847 799 D
1945 668 D
2043 583 D
2141 521 D
2240 470 D
2338 432 D
2436 403 D
2534 375 D
2632 355 D
2730 336 D
2828 323 D
2926 314 D
3025 302 D
3123 295 D
3221 290 D
3319 284 D
3417 284 D
LT0
3114 1746 M
180 0 V
1642 1166 M
29 -102 V
35 -87 V
71 -132 V
64 -86 V
56 -63 V
51 -48 V
47 -39 V
43 -32 V
40 -26 V
37 -22 V
34 -19 V
32 -17 V
31 -15 V
29 -13 V
27 -11 V
26 -11 V
25 -9 V
24 -8 V
23 -8 V
21 -7 V
21 -7 V
stroke
grestore
end
showpage
}
\put(3054,1746){\makebox(0,0)[r]{variational}}
\put(3054,1846){\makebox(0,0)[r]{Monte Carlo}}
\put(2008,2109){\makebox(0,0){Internal energy, $XY$ model, $64\times 64$}}
\put(2008,51){\makebox(0,0){$\beta$}}
\put(100,1130){%
\special{ps: gsave currentpoint currentpoint translate
270 rotate neg exch neg exch translate}%
\makebox(0,0)[b]{\shortstack{$U$}}%
\special{ps: currentpoint grestore moveto}%
}
\put(3048,151){\makebox(0,0){10}}
\put(1824,151){\makebox(0,0){1}}
\put(600,151){\makebox(0,0){0.1}}
\put(540,2009){\makebox(0,0)[r]{2}}
\put(540,1833){\makebox(0,0)[r]{1.8}}
\put(540,1657){\makebox(0,0)[r]{1.6}}
\put(540,1482){\makebox(0,0)[r]{1.4}}
\put(540,1306){\makebox(0,0)[r]{1.2}}
\put(540,1130){\makebox(0,0)[r]{1}}
\put(540,954){\makebox(0,0)[r]{0.8}}
\put(540,778){\makebox(0,0)[r]{0.6}}
\put(540,603){\makebox(0,0)[r]{0.4}}
\put(540,427){\makebox(0,0)[r]{0.2}}
\put(540,251){\makebox(0,0)[r]{0}}
\end{picture}
\caption{Internal energy per site of the two-dimensional $XY$ model on
$64\times 64$ lattice, calculated variationally with wavelets and by a
Monte Carlo method.}
\label{fig4}
\end{figure}

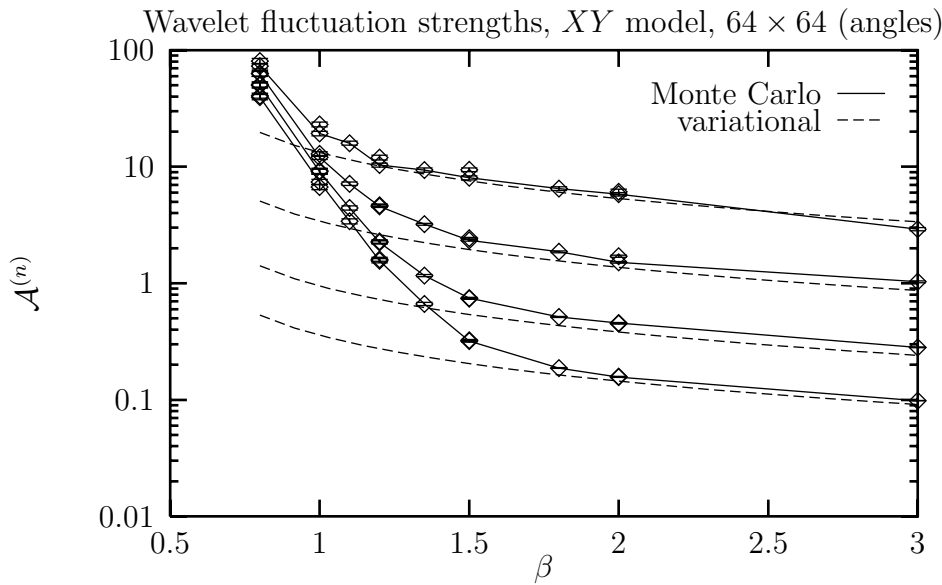
\begin{figure}
\setlength{\unitlength}{0.1bp}
\special{!
/gnudict 40 dict def
gnudict begin
/Color false def
/Solid false def
/gnulinewidth 5.000 def
/vshift -33 def
/dl {10 mul} def
/hpt 31.5 def
/vpt 31.5 def
/M {moveto} bind def
/L {lineto} bind def
/R {rmoveto} bind def
/V {rlineto} bind def
/vpt2 vpt 2 mul def
/hpt2 hpt 2 mul def
/Lshow { currentpoint stroke M
  0 vshift R show } def
/Rshow { currentpoint stroke M
  dup stringwidth pop neg vshift R show } def
/Cshow { currentpoint stroke M
  dup stringwidth pop -2 div vshift R show } def
/DL { Color {setrgbcolor Solid {pop []} if 0 setdash }
 {pop pop pop Solid {pop []} if 0 setdash} ifelse } def
/BL { stroke gnulinewidth 2 mul setlinewidth } def
/AL { stroke gnulinewidth 2 div setlinewidth } def
/PL { stroke gnulinewidth setlinewidth } def
/LTb { BL [] 0 0 0 DL } def
/LTa { AL [1 dl 2 dl] 0 setdash 0 0 0 setrgbcolor } def
/LT0 { PL [] 0 1 0 DL } def
/LT1 { PL [4 dl 2 dl] 0 0 1 DL } def
/LT2 { PL [2 dl 3 dl] 1 0 0 DL } def
/LT3 { PL [1 dl 1.5 dl] 1 0 1 DL } def
/LT4 { PL [5 dl 2 dl 1 dl 2 dl] 0 1 1 DL } def
/LT5 { PL [4 dl 3 dl 1 dl 3 dl] 1 1 0 DL } def
/LT6 { PL [2 dl 2 dl 2 dl 4 dl] 0 0 0 DL } def
/LT7 { PL [2 dl 2 dl 2 dl 2 dl 2 dl 4 dl] 1 0.3 0 DL } def
/LT8 { PL [2 dl 2 dl 2 dl 2 dl 2 dl 2 dl 2 dl 4 dl] 0.5 0.5 0.5 DL } def
/P { stroke [] 0 setdash
  currentlinewidth 2 div sub M
  0 currentlinewidth V stroke } def
/D { stroke [] 0 setdash 2 copy vpt add M
  hpt neg vpt neg V hpt vpt neg V
  hpt vpt V hpt neg vpt V closepath stroke
  P } def
/A { stroke [] 0 setdash vpt sub M 0 vpt2 V
  currentpoint stroke M
  hpt neg vpt neg R hpt2 0 V stroke
  } def
/B { stroke [] 0 setdash 2 copy exch hpt sub exch vpt add M
  0 vpt2 neg V hpt2 0 V 0 vpt2 V
  hpt2 neg 0 V closepath stroke
  P } def
/C { stroke [] 0 setdash exch hpt sub exch vpt add M
  hpt2 vpt2 neg V currentpoint stroke M
  hpt2 neg 0 R hpt2 vpt2 V stroke } def
/T { stroke [] 0 setdash 2 copy vpt 1.12 mul add M
  hpt neg vpt -1.62 mul V
  hpt 2 mul 0 V
  hpt neg vpt 1.62 mul V closepath stroke
  P  } def
/S { 2 copy A C} def
end
}
\begin{picture}(3600,2160)(0,0)
\special{"
gnudict begin
gsave
50 50 translate
0.100 0.100 scale
0 setgray
/Helvetica findfont 100 scalefont setfont
newpath
-500.000000 -500.000000 translate
LTa
LTb
600 251 M
63 0 V
2754 0 R
-63 0 V
600 383 M
31 0 V
2786 0 R
-31 0 V
600 461 M
31 0 V
2786 0 R
-31 0 V
600 516 M
31 0 V
2786 0 R
-31 0 V
600 558 M
31 0 V
2786 0 R
-31 0 V
600 593 M
31 0 V
2786 0 R
-31 0 V
600 622 M
31 0 V
2786 0 R
-31 0 V
600 648 M
31 0 V
2786 0 R
-31 0 V
600 670 M
31 0 V
2786 0 R
-31 0 V
600 690 M
63 0 V
2754 0 R
-63 0 V
600 823 M
31 0 V
2786 0 R
-31 0 V
600 900 M
31 0 V
2786 0 R
-31 0 V
600 955 M
31 0 V
2786 0 R
-31 0 V
600 998 M
31 0 V
2786 0 R
-31 0 V
600 1032 M
31 0 V
2786 0 R
-31 0 V
600 1062 M
31 0 V
2786 0 R
-31 0 V
600 1087 M
31 0 V
2786 0 R
-31 0 V
600 1110 M
31 0 V
2786 0 R
-31 0 V
600 1130 M
63 0 V
2754 0 R
-63 0 V
600 1262 M
31 0 V
2786 0 R
-31 0 V
600 1340 M
31 0 V
2786 0 R
-31 0 V
600 1395 M
31 0 V
2786 0 R
-31 0 V
600 1437 M
31 0 V
2786 0 R
-31 0 V
600 1472 M
31 0 V
2786 0 R
-31 0 V
600 1501 M
31 0 V
2786 0 R
-31 0 V
600 1527 M
31 0 V
2786 0 R
-31 0 V
600 1549 M
31 0 V
2786 0 R
-31 0 V
600 1570 M
63 0 V
2754 0 R
-63 0 V
600 1702 M
31 0 V
2786 0 R
-31 0 V
600 1779 M
31 0 V
2786 0 R
-31 0 V
600 1834 M
31 0 V
2786 0 R
-31 0 V
600 1877 M
31 0 V
2786 0 R
-31 0 V
600 1911 M
31 0 V
2786 0 R
-31 0 V
600 1941 M
31 0 V
2786 0 R
-31 0 V
600 1966 M
31 0 V
2786 0 R
-31 0 V
600 1989 M
31 0 V
2786 0 R
-31 0 V
600 2009 M
63 0 V
2754 0 R
-63 0 V
600 251 M
0 63 V
0 1695 R
0 -63 V
1163 251 M
0 63 V
0 1695 R
0 -63 V
1727 251 M
0 63 V
0 1695 R
0 -63 V
2290 251 M
0 63 V
0 1695 R
0 -63 V
2854 251 M
0 63 V
0 1695 R
0 -63 V
3417 251 M
0 63 V
0 1695 R
0 -63 V
600 251 M
2817 0 V
0 1758 V
-2817 0 V
600 251 L
LT0
3114 1846 M
180 0 V
938 1832 M
225 -320 V
113 -148 V
113 -149 V
169 -163 V
1727 912 L
2065 810 L
225 -34 V
3417 687 L
938 1837 M
225 -344 R
226 -273 R
1727 915 M
2290 778 M
938 1876 M
225 -325 V
113 -137 V
113 -130 V
169 -125 V
169 -85 V
338 -71 V
225 -23 V
3417 888 L
938 1880 M
225 -326 R
226 -266 R
338 -216 R
563 -94 R
938 1920 M
225 -315 V
113 -100 V
113 -85 V
169 -67 V
169 -61 V
338 -43 V
225 -40 V
1127 -73 V
938 1922 M
225 -304 R
226 -194 R
338 -124 R
563 -68 R
938 1949 M
225 -254 V
113 -37 V
113 -82 V
169 -19 V
169 -29 V
338 -41 V
225 -21 V
3417 1334 L
938 1968 M
225 -239 R
226 -125 R
338 -46 R
563 -83 R
938 1832 D
1163 1512 D
1276 1364 D
1389 1215 D
1558 1052 D
1727 912 D
2065 810 D
2290 776 D
3417 687 D
938 1837 D
1163 1493 D
1389 1220 D
1727 915 D
2290 778 D
938 1876 D
1163 1551 D
1276 1414 D
1389 1284 D
1558 1159 D
1727 1074 D
2065 1003 D
2290 980 D
3417 888 D
938 1880 D
1163 1554 D
1389 1288 D
1727 1072 D
2290 978 D
938 1920 D
1163 1605 D
1276 1505 D
1389 1420 D
1558 1353 D
1727 1292 D
2065 1249 D
2290 1209 D
3417 1136 D
938 1922 D
1163 1618 D
1389 1424 D
1727 1300 D
2290 1232 D
938 1949 D
1163 1695 D
1276 1658 D
1389 1576 D
1558 1557 D
1727 1528 D
2065 1487 D
2290 1466 D
3417 1334 D
938 1968 D
1163 1729 D
1389 1604 D
1727 1558 D
2290 1475 D
938 1822 M
0 20 V
-31 -20 R
62 0 V
-62 20 R
62 0 V
194 -340 R
0 20 V
-31 -20 R
62 0 V
-62 20 R
62 0 V
82 -168 R
0 19 V
-31 -19 R
62 0 V
-62 19 R
62 0 V
82 -166 R
0 16 V
-31 -16 R
62 0 V
-62 16 R
62 0 V
138 -177 R
0 12 V
-31 -12 R
62 0 V
-62 12 R
62 0 V
1727 908 M
0 7 V
-31 -7 R
62 0 V
-62 7 R
62 0 V
2065 809 M
0 3 V
-31 -3 R
62 0 V
-62 3 R
62 0 V
194 -37 R
0 2 V
-31 -2 R
62 0 V
-62 2 R
62 0 V
3417 687 M
0 1 V
-31 -1 R
62 0 V
-62 1 R
62 0 V
938 1828 M
0 19 V
-31 -19 R
62 0 V
-62 19 R
62 0 V
194 -364 R
0 20 V
-31 -20 R
62 0 V
-62 20 R
62 0 V
195 -290 R
0 15 V
-31 -15 R
62 0 V
-62 15 R
62 0 V
1727 911 M
0 7 V
-31 -7 R
62 0 V
-62 7 R
62 0 V
2290 777 M
0 2 V
-31 -2 R
62 0 V
-62 2 R
62 0 V
938 1866 M
0 19 V
-31 -19 R
62 0 V
-62 19 R
62 0 V
194 -344 R
0 19 V
-31 -19 R
62 0 V
-62 19 R
62 0 V
82 -154 R
0 16 V
-31 -16 R
62 0 V
-62 16 R
62 0 V
82 -144 R
0 12 V
-31 -12 R
62 0 V
-62 12 R
62 0 V
138 -135 R
0 8 V
-31 -8 R
62 0 V
-62 8 R
62 0 V
138 -91 R
0 5 V
-31 -5 R
62 0 V
-62 5 R
62 0 V
307 -76 R
0 4 V
-31 -4 R
62 0 V
-62 4 R
62 0 V
194 -27 R
0 4 V
-31 -4 R
62 0 V
-62 4 R
62 0 V
3417 887 M
0 3 V
-31 -3 R
62 0 V
-62 3 R
62 0 V
938 1870 M
0 20 V
-31 -20 R
62 0 V
-62 20 R
62 0 V
194 -346 R
0 20 V
-31 -20 R
62 0 V
-62 20 R
62 0 V
195 -283 R
0 13 V
-31 -13 R
62 0 V
-62 13 R
62 0 V
307 -224 R
0 4 V
-31 -4 R
62 0 V
-62 4 R
62 0 V
532 -97 R
0 3 V
-31 -3 R
62 0 V
-62 3 R
62 0 V
938 1910 M
0 19 V
-31 -19 R
62 0 V
-62 19 R
62 0 V
194 -332 R
0 15 V
-31 -15 R
62 0 V
-62 15 R
62 0 V
82 -113 R
0 13 V
-31 -13 R
62 0 V
-62 13 R
62 0 V
82 -96 R
0 9 V
-31 -9 R
62 0 V
-62 9 R
62 0 V
138 -76 R
0 7 V
-31 -7 R
62 0 V
-62 7 R
62 0 V
138 -67 R
0 7 V
-31 -7 R
62 0 V
-62 7 R
62 0 V
307 -51 R
0 7 V
-31 -7 R
62 0 V
-62 7 R
62 0 V
194 -46 R
0 6 V
-31 -6 R
62 0 V
-62 6 R
62 0 V
1096 -80 R
0 7 V
-31 -7 R
62 0 V
-62 7 R
62 0 V
938 1911 M
0 20 V
-31 -20 R
62 0 V
-62 20 R
62 0 V
194 -322 R
0 17 V
-31 -17 R
62 0 V
-62 17 R
62 0 V
195 -207 R
0 10 V
-31 -10 R
62 0 V
-62 10 R
62 0 V
307 -133 R
0 7 V
-31 -7 R
62 0 V
-62 7 R
62 0 V
532 -75 R
0 8 V
-31 -8 R
62 0 V
-62 8 R
62 0 V
938 1938 M
0 21 V
-31 -21 R
62 0 V
-62 21 R
62 0 V
194 -273 R
0 17 V
-31 -17 R
62 0 V
-62 17 R
62 0 V
82 -52 R
0 15 V
-31 -15 R
62 0 V
-62 15 R
62 0 V
82 -98 R
0 15 V
-31 -15 R
62 0 V
-62 15 R
62 0 V
138 -34 R
0 16 V
-31 -16 R
62 0 V
-62 16 R
62 0 V
138 -45 R
0 15 V
-31 -15 R
62 0 V
-62 15 R
62 0 V
307 -55 R
0 14 V
-31 -14 R
62 0 V
-62 14 R
62 0 V
194 -35 R
0 13 V
-31 -13 R
62 0 V
-62 13 R
62 0 V
3417 1328 M
0 12 V
-31 -12 R
62 0 V
-62 12 R
62 0 V
938 1957 M
0 21 V
-31 -21 R
62 0 V
-62 21 R
62 0 V
194 -258 R
0 18 V
-31 -18 R
62 0 V
-62 18 R
62 0 V
195 -143 R
0 18 V
-31 -18 R
62 0 V
-62 18 R
62 0 V
307 -62 R
0 14 V
-31 -14 R
62 0 V
-62 14 R
62 0 V
532 -100 R
0 20 V
-31 -20 R
62 0 V
-62 20 R
62 0 V
LT1
3114 1746 M
180 0 V
938 1010 M
131 -49 V
130 -36 V
130 -30 V
131 -25 V
130 -22 V
131 -20 V
130 -18 V
131 -16 V
130 -15 V
131 -13 V
130 -13 V
131 -12 V
130 -12 V
131 -10 V
130 -10 V
131 -10 V
130 -9 V
131 -9 V
130 -8 V
938 1196 M
131 -49 V
130 -37 V
130 -29 V
131 -26 V
130 -22 V
131 -20 V
130 -17 V
131 -16 V
130 -15 V
131 -14 V
130 -13 V
131 -12 V
130 -11 V
131 -11 V
130 -10 V
131 -9 V
130 -10 V
131 -8 V
130 -9 V
938 1440 M
131 -49 V
130 -36 V
130 -30 V
131 -25 V
130 -22 V
131 -20 V
130 -18 V
131 -16 V
130 -15 V
131 -14 V
130 -13 V
131 -12 V
130 -11 V
131 -11 V
130 -10 V
131 -9 V
130 -9 V
131 -9 V
130 -8 V
938 1699 M
131 -49 V
130 -36 V
130 -30 V
131 -25 V
130 -22 V
131 -20 V
130 -18 V
131 -16 V
130 -15 V
131 -14 V
130 -12 V
131 -12 V
130 -11 V
131 -11 V
130 -10 V
131 -10 V
130 -9 V
131 -9 V
130 -8 V
stroke
grestore
end
showpage
}
\put(3054,1746){\makebox(0,0)[r]{variational}}
\put(3054,1846){\makebox(0,0)[r]{Monte Carlo}}
\put(2008,2109){\makebox(0,0){Wavelet fluctuation strengths, $XY$ model, $64\times 64$ (angles)}}
\put(2008,51){\makebox(0,0){$\beta$}}
\put(100,1130){%
\special{ps: gsave currentpoint currentpoint translate
270 rotate neg exch neg exch translate}%
\makebox(0,0)[b]{\shortstack{${\cal A}^{(n)}$}}%
\special{ps: currentpoint grestore moveto}%
}
\put(3417,151){\makebox(0,0){3}}
\put(2854,151){\makebox(0,0){2.5}}
\put(2290,151){\makebox(0,0){2}}
\put(1727,151){\makebox(0,0){1.5}}
\put(1163,151){\makebox(0,0){1}}
\put(600,151){\makebox(0,0){0.5}}
\put(540,2009){\makebox(0,0)[r]{100}}
\put(540,1570){\makebox(0,0)[r]{10}}
\put(540,1130){\makebox(0,0)[r]{1}}
\put(540,690){\makebox(0,0)[r]{0.1}}
\put(540,251){\makebox(0,0)[r]{0.01}}
\end{picture}
\caption{Wavelet coefficient fluctuations in the two-dimensional $XY$ model
on a $64\times 64$ lattice. \label{fig5}}
\end{figure}

Above $\beta\approx 1.5$, variational and Monte Carlo result coincide.
Below that, nongaussian fluctuations due to periodicity begin to play a
role, and the variational method fails. This is also manifest when $\langle
[a^{(n)}_t(x')]^4 \rangle$ is measured where a strong departure from
Gaussian factorization can be observed.

\section{Landau-Ginzburg model}

In the example of the $XY$ model, major problems were caused by having the
(periodic) group parameter (i.e., the angle) as field variable instead of
the spin vectors. Since the constraint $s^2 = 1$ on the length of the spin
vectors translates to a complicated expression in wavelet space, it is not
easily incorporated in the variational ansatz. (The same problem prevented
us from considering the even more simple Ising model.)

However, as our main interest will be in renormalization group
transformations which dilute the constraints and eventually allow a
description in terms of a continuous order parameter governed by an effective
theory, we now turn to the study of this effective theory, the
Landau-Ginzburg model.

\subsection{Wavelet representation}

The Landau-Ginzburg model is given by a $N_c$-component field $a(x)$ with
the $O(N_c)$-symmetric Hamiltonian
\be \label{Hxy}
  {\cal H} = \frac{1}{2} \sum_{x\in{\cal L}} \sum_{\mu=1}^n
             \left( a(x+e_\mu) - a(x) \right)
            +\frac{m^2}{2} \sum_{x\in{\cal L}} a(x)^2
            +\frac{g}{2}  \sum_{x\in{\cal L}} \left(a(x)^2\right)^2
\ee
corresponding to a $(\Phi^2)^2$ field theory. For $m^2<0$ and $g>0$ it
exhibits a ``Mexican hat'' potential with spontaneous symmetry breaking.
In mean-field theory, the phase transition between $O(N_c)$ symmetric and
symmetry-broken phase takes place at $\beta=0$. We wish to estimate the
shift of the phase transition temperature due to fluctuations in wavelets.

To allow for symmetry breaking, we must add another variational parameter,
the expectation value $\bar a$ of the field:
\be
  \langle a(x) \rangle = \bar a \quad.
\ee
As this is a constant, it implies for the wavelet coefficients
\be
  \langle \hat a^{(n)}_{t}(x') \rangle =
  \delta_{n,N} \delta_{t,0} \frac{\bar a}{\sqrt{N_0}} \quad,
\ee
i.e., the only coefficient with a nonzero expectation value is $\hat
a^{(N)}_0$, associated with the scaling function at the topmost level. The
additional variational parameter $\bar a$, giving the center of the
probability distribution of $\hat a^{(N)}_0$, appears only in the internal
energy, as a mere shift of the probability distribution does not influence
the entropy.

The first two terms of the Hamiltonian (kinetic and mass) have been
calculated in the context of the Gaussian model. The only modification is
that we now have to deal with a vector-valued field $a_\alpha(x)$,
$\alpha=1,\ldots,N_c$. The wavelet fluctuation strength is thus a tensor
\be
  \langle \hat a^{(n_1)}_{t_1,\alpha_1}(x'_1) \,
          \hat a^{(n_2)}_{t_2,\alpha_2}(x'_2) \rangle
  = \delta_{n_1,n_2} \delta_{t_1,t_2} \delta_{x'_1,x'_2}
    \, {\cal A}^{(n_1)}_{t_1,\alpha_1\alpha_2} \quad.
\ee
The unperturbed Hamiltonian appears then as
\be
  {\cal H}_0 = \frac{1}{2} \sum_{n=1}^N \sum_t^{n_w}
               |L^{(n)}_t|^2 \, \tr {\cal A}^{(n)}_{t}
             + \frac{m^2}{2} \sum_{n=1}^N \sum_t^{n_w}
               N_n \,\tr{\cal A}^{(n)}_t
             + \frac{m^2}{2} N_0 \, \bar a^2
\ee
where the trace operator refers to summing over the vector indices
$\alpha$.

To calculate the influence of interactions, we expand the third term of
(\ref{Hxy}) in wavelet space and then can make use of Wick's theorem, as our
trial ensemble is Gaussian:
\be
  \langle a b c d \rangle = \langle a b \rangle \, \langle c d \rangle
  + \langle a c \rangle \, \langle b d \rangle
  + \langle a d \rangle \, \langle b c \rangle \quad.
\ee
Then
\bea &&
  \frac{g}{2} \sum_x \left(\sum_\alpha a_\alpha(x)^2\right)^2
  = \sum_x \sum_{\alpha\beta}
    a_\alpha(x) a_\alpha(x) a_\beta(x) a_\beta(x) \nnm\\
  && \qquad = \frac{g}{2}
  \sum_{n_1,n_2}^N \sum_{t_1,t_2}^{n_w} \sum_{x'_1,x'_2} \sum_x
    \left[\psi^{(n_1)}_{t_1}(x'_1)(x)\right]^2 \,
    \left[\psi^{(n_2)}_{t_2}(x'_2)(x)\right]^2 \, \nnm\\ &&\qquad\qquad\qquad
  \times
    \left(\tr{\cal A}^{n_1}_{t_1}\,\tr{\cal A}^{n_2}_{t_2}
        +2\tr({\cal A}^{n_1}_{t_1}\,{\cal A}^{n_2}_{t_2})\right)
  \nnm\\ &&\qquad\qquad + g
  \sum_n^N \sum_t^{n_w} \sum_{x'} \sum_x \left[\psi^{(n)}_t(x')(x)\right]^2 \,
    \bar a^2
  \, \left( \tr{\cal A}^{(n)}_{t}
                 +2 {\cal A}^{(n)}_{t,11} \right) \nnm\\ && \qquad\qquad
  + \frac{g}{2} N_0 \bar a^4  \quad.
\eea
We take the symmetry-breaking along the $1$-direction in $O(N_c)$ space. The
fluctuation matrix at each scale retains the symmetry under the corresponding
stabilizer group (i.e.~the group of transformations that leave the symmetry
breaking vector invariant) and is thus of the form
\be
  {\cal A} = \left(\begin{array}{cccc}
                   {\cal A}_{\par} & 0 & 0 & \cdots \\
                   0 & {\cal A}_{\perp} & 0 & \cdots \\
                   0 & 0 & {\cal A}_{\perp} & \cdots \\
                   \vdots & \vdots & \vdots & \ddots
                   \end{array}\right) \quad.
\ee
${\cal A}_\prll$ gives the fluctuation strength in the direction of the
symmetry breaking, ${\cal A}_\perp$ perpendicular to it. In particular
\bea
  \tr{\cal A} &=& {\cal A}_\prll + (N_c-1) {\cal A}_\perp \\
  \tr({\cal A}{\cal A}') &=&
  {\cal A}_\prll {\cal A}'_\prll + (N_c-1) {\cal A}_\perp {\cal A}'_\perp
  \quad.
\eea

With this, the complete internal energy is
\bea\label{phi4}
  U &=& \frac{1}{2} \sum_{n=1}^N \sum_t^{n_w}
              \left( |L^{(n)}_t|^2 + m^2 N_n \right)
        \left( {\cal A}^{(n)}_{\prll,t} + (N_c-1) {\cal A}^{(n)}_{\perp,t}
                                                                     \right)
  \nnm\\ &+& \frac{g}{2}
  \sum_{n_1n_2}^N \sum_{t_1t_2}^{n_w} M^{n_1n_2}_{t_1t_2} \,
  \Big[ 2 {\cal A}^{(n_1)}_{\prll,t_1} {\cal A}^{(n_2)}_{\prll,t_2}
        + (N_c^2-1) {\cal A}^{(n_1)}_{\perp,t_1} {\cal A}^{(n_2)}_{\perp,t_2}
  \nnm\\&&\qquad\qquad\qquad\qquad
        + 2 (N_c-1) {\cal A}^{(n_1)}_{\prll,t_1} {\cal A}^{(n_2)}_{\perp,t_2}
  \Big]
  \nnm\\ &+& g \, \bar a^2
  \sum_n^N \sum_t^{n_w} N_n \,
  \left( 3 {\cal A}^{(n)}_{\prll,t} + (N_c-1) {\cal A}^{(n)}_{\perp,t} \right)
  + \frac{m^2}{2} N_0 \bar a^2 + \frac{g}{2} N_0 \bar a^4 \quad.
\eea
The first term gives the original Gaussian Hamiltonian, the second
describes interactions between different scales due to the $\Phi^4$
interaction term; the final two terms are due to spontaneous symmetry
breaking.

We have introduced the matrix
\be
  M^{n_1 n_2}_{t_1 t_2} =
  \sum_{x\in{\cal L}} \sum_{x'_1\in{\cal L}^1} \sum_{x'_2\in{\cal L}^2}
  \left[ \psi^{(n_1)}_{t_1}(x'_1)(x) \right]^2 \,
  \left[ \psi^{(n_2)}_{t_2}(x'_2)(x) \right]^2
\ee
describing the nonlinear interaction of fluctuations at different scales
caused by the quartic term. Due to the factorization properties of wavelets,
it can be rewritten as
\be
  M^{n_1 n_2}_{t_1 t_2} = \prod_{k=1}^D {\cal M}^{n_1 n_2}_{t_{1k} t_{2k}}
\ee
with
\be
  {\cal M}^{n_1 n_2}_{t_1 t_2} = N_{n_1}
  \sum_{x'_2\in{\cal L}^{n_2}} \sum_{x\in{\cal L}}
  \left[ \psi^{(1D,n_1)}_{t_1}(x-x'_1) \right]^2 \,
  \left[ \psi^{(1D,n_2)}_{t_2}(x-x'_2) \right]^2
\ee
derived from one-dimensional wavelets. In particular, the complexity of
(\ref{phi4}) increases only slightly when higher dimensions are considered.

\subsection{Spontaneous symmetry breaking}

The value of $\bar a$ can be calculated analytically. As it does not enter
the entropy, it is determined by the minimum of the internal energy if the
${\cal A}^{(n)}_t$ are given. The corresponding condition is
\be
  \frac{\partial F}{\partial \bar a} =
  m^2 N_0 \bar a +
  2 N_0 g \bar a^3 +
  2 g \bar a \sum_n^N \sum_t^{n_w} N_n
      \left( 3 {\cal A}^{(n)}_{\prll,t} +
             (N_c-1) {\cal A}^{(n)}_{\perp,t} \right)
  = 0 \quad.
\ee
While $\bar a = 0$ is always one solution to this equation, the other one is
the root of a quadratic equation,
\be \label{abar}
  \bar a = \sqrt{ - \frac{m^2}{2g}
    - \sum_n^N \sum_t^{n_w} N_n
      \left( 3 {\cal A}^{(n)}_{\prll,t} +
             (N_c-1) {\cal A}^{(n)}_{\perp,t} \right)} \quad.
\ee
Without the fluctuations, this is just the mean field result which exhibits
spontaneous symmetry breaking for all values of $\beta$, provided that $m^2/2g
< 0$. The more fluctuations appear, the more $\bar a$ is reduced, until the
expression under the root becomes negative, and $\bar a = 0$ becomes the
only solution, corresponding to a symmetric phase.

The value of $\bar a^2$ interacts with fluctuations due to the third term
in (\ref{phi4}). As the fluctuations enter linearly and indiscriminately of
their scale, this modifies the effective mass of the theory. Due to the
prefactor $3$ in the fluctuation sum, it is different for parallel and for
perpendicular fluctuations.  Inserting the mean-field value of $\bar a$ for
comparison, we find
\be
  m \longrightarrow \left\{\begin{array}{l}
    m_{\prll} = m + 6g\bar a^2 \approx -2m \\
    m_{\perp} = m + 2g\bar a^2 \approx 0
    \end{array}\right.
\ee
The value $-2m$ is the correct Gaussian curvature of the potential for
fluctuations against the valley. Fluctuations perpendicular to the
magnetization are massless Goldstone modes.

\subsection{Numerical solution}

We tested the accuracy of this model both for the case of a small anharmonic
perturbation and the full phase transition. Eq.~(\ref{phi4}) was minimized
numerically using a combination of the simplex method and simulated
annealing \cite{NR}.

For comparison, a simple Metropolis simulation of the model was performed.
The wavelet transform was calculated on statistically independent
configurations and the fluctuation strength of wavelet coefficients measured
in this ensemble.

One should keep in mind that the trial probability distribution used here
makes at least two bold approximations: It neglects all correlations between
wavelet fluctuations, and it assumes that all fluctuations are Gaussian. We
therefore should expect no more than a gross representation of the model.
However, this should be sufficient for a wavelet renormalization group to
work, as the renormalization transformations should enhance the physically
important features.

\subsubsection{Small anharmonic perturbations}

\begin{figure}
\setlength{\unitlength}{0.1bp}
\special{!
/gnudict 40 dict def
gnudict begin
/Color false def
/Solid false def
/gnulinewidth 5.000 def
/vshift -33 def
/dl {10 mul} def
/hpt 31.5 def
/vpt 31.5 def
/M {moveto} bind def
/L {lineto} bind def
/R {rmoveto} bind def
/V {rlineto} bind def
/vpt2 vpt 2 mul def
/hpt2 hpt 2 mul def
/Lshow { currentpoint stroke M
  0 vshift R show } def
/Rshow { currentpoint stroke M
  dup stringwidth pop neg vshift R show } def
/Cshow { currentpoint stroke M
  dup stringwidth pop -2 div vshift R show } def
/DL { Color {setrgbcolor Solid {pop []} if 0 setdash }
 {pop pop pop Solid {pop []} if 0 setdash} ifelse } def
/BL { stroke gnulinewidth 2 mul setlinewidth } def
/AL { stroke gnulinewidth 2 div setlinewidth } def
/PL { stroke gnulinewidth setlinewidth } def
/LTb { BL [] 0 0 0 DL } def
/LTa { AL [1 dl 2 dl] 0 setdash 0 0 0 setrgbcolor } def
/LT0 { PL [] 0 1 0 DL } def
/LT1 { PL [4 dl 2 dl] 0 0 1 DL } def
/LT2 { PL [2 dl 3 dl] 1 0 0 DL } def
/LT3 { PL [1 dl 1.5 dl] 1 0 1 DL } def
/LT4 { PL [5 dl 2 dl 1 dl 2 dl] 0 1 1 DL } def
/LT5 { PL [4 dl 3 dl 1 dl 3 dl] 1 1 0 DL } def
/LT6 { PL [2 dl 2 dl 2 dl 4 dl] 0 0 0 DL } def
/LT7 { PL [2 dl 2 dl 2 dl 2 dl 2 dl 4 dl] 1 0.3 0 DL } def
/LT8 { PL [2 dl 2 dl 2 dl 2 dl 2 dl 2 dl 2 dl 4 dl] 0.5 0.5 0.5 DL } def
/P { stroke [] 0 setdash
  currentlinewidth 2 div sub M
  0 currentlinewidth V stroke } def
/D { stroke [] 0 setdash 2 copy vpt add M
  hpt neg vpt neg V hpt vpt neg V
  hpt vpt V hpt neg vpt V closepath stroke
  P } def
/A { stroke [] 0 setdash vpt sub M 0 vpt2 V
  currentpoint stroke M
  hpt neg vpt neg R hpt2 0 V stroke
  } def
/B { stroke [] 0 setdash 2 copy exch hpt sub exch vpt add M
  0 vpt2 neg V hpt2 0 V 0 vpt2 V
  hpt2 neg 0 V closepath stroke
  P } def
/C { stroke [] 0 setdash exch hpt sub exch vpt add M
  hpt2 vpt2 neg V currentpoint stroke M
  hpt2 neg 0 R hpt2 vpt2 V stroke } def
/T { stroke [] 0 setdash 2 copy vpt 1.12 mul add M
  hpt neg vpt -1.62 mul V
  hpt 2 mul 0 V
  hpt neg vpt 1.62 mul V closepath stroke
  P  } def
/S { 2 copy A C} def
end
}
\begin{picture}(3600,2160)(0,0)
\special{"
gnudict begin
gsave
50 50 translate
0.100 0.100 scale
0 setgray
/Helvetica findfont 100 scalefont setfont
newpath
-500.000000 -500.000000 translate
LTa
LTb
600 251 M
63 0 V
2754 0 R
-63 0 V
600 581 M
31 0 V
2786 0 R
-31 0 V
600 775 M
31 0 V
2786 0 R
-31 0 V
600 912 M
31 0 V
2786 0 R
-31 0 V
600 1018 M
31 0 V
2786 0 R
-31 0 V
600 1105 M
31 0 V
2786 0 R
-31 0 V
600 1178 M
31 0 V
2786 0 R
-31 0 V
600 1242 M
31 0 V
2786 0 R
-31 0 V
600 1298 M
31 0 V
2786 0 R
-31 0 V
600 1348 M
63 0 V
2754 0 R
-63 0 V
600 1679 M
31 0 V
2786 0 R
-31 0 V
600 1872 M
31 0 V
2786 0 R
-31 0 V
600 2009 M
31 0 V
2786 0 R
-31 0 V
600 251 M
0 63 V
0 1695 R
0 -63 V
913 251 M
0 63 V
0 1695 R
0 -63 V
1226 251 M
0 63 V
0 1695 R
0 -63 V
1539 251 M
0 63 V
0 1695 R
0 -63 V
1852 251 M
0 63 V
0 1695 R
0 -63 V
2165 251 M
0 63 V
0 1695 R
0 -63 V
2478 251 M
0 63 V
0 1695 R
0 -63 V
2791 251 M
0 63 V
0 1695 R
0 -63 V
3104 251 M
0 63 V
0 1695 R
0 -63 V
3417 251 M
0 63 V
0 1695 R
0 -63 V
600 251 M
2817 0 V
0 1758 V
-2817 0 V
600 251 L
LT0
3114 1846 M
180 0 V
600 1501 M
748 1356 L
897 1247 L
148 -87 V
148 -73 V
148 -62 V
149 -56 V
148 -49 V
148 -44 V
148 -41 V
149 -37 V
148 -35 V
148 -32 V
148 -31 V
149 -28 V
148 -27 V
148 -25 V
148 -24 V
149 -23 V
148 -22 V
600 1719 M
748 1590 L
149 -97 V
148 -79 V
148 -67 V
148 -57 V
149 -51 V
148 -46 V
148 -41 V
148 -38 V
149 -35 V
148 -33 V
148 -31 V
148 -27 V
149 -28 V
148 -25 V
148 -24 V
148 -23 V
149 -22 V
148 -21 V
600 1872 M
748 1760 L
149 -84 V
148 -71 V
148 -58 V
148 -51 V
149 -45 V
148 -42 V
148 -38 V
148 -34 V
149 -33 V
148 -30 V
148 -28 V
148 -26 V
149 -25 V
148 -24 V
148 -22 V
148 -22 V
149 -20 V
148 -20 V
600 1928 M
748 1825 L
149 -77 V
148 -65 V
148 -54 V
148 -48 V
149 -43 V
148 -39 V
148 -35 V
148 -33 V
149 -30 V
148 -29 V
148 -26 V
148 -26 V
149 -25 V
148 -20 V
148 -23 V
148 -21 V
149 -19 V
148 -19 V
LT1
3114 1746 M
180 0 V
600 1547 M
913 1273 L
313 -166 V
1539 982 L
313 -95 V
313 -83 V
313 -69 V
313 -57 V
313 -53 V
313 -48 V
600 1763 M
913 1522 L
313 -154 V
313 -110 V
313 -94 V
313 -77 V
313 -54 V
313 -65 V
313 -50 V
313 -41 V
600 1899 M
913 1690 L
313 -131 V
313 -100 V
313 -83 V
313 -69 V
313 -63 V
313 -53 V
313 -52 V
313 -47 V
600 1949 M
913 1727 L
313 -108 V
313 -80 V
313 -81 V
313 -81 V
313 -60 V
313 2 V
313 -73 V
313 -38 V
3174 1746 D
600 1547 D
913 1273 D
1226 1107 D
1539 982 D
1852 887 D
2165 804 D
2478 735 D
2791 678 D
3104 625 D
3417 577 D
600 1763 D
913 1522 D
1226 1368 D
1539 1258 D
1852 1164 D
2165 1087 D
2478 1033 D
2791 968 D
3104 918 D
3417 877 D
600 1899 D
913 1690 D
1226 1559 D
1539 1459 D
1852 1376 D
2165 1307 D
2478 1244 D
2791 1191 D
3104 1139 D
3417 1092 D
600 1949 D
913 1727 D
1226 1619 D
1539 1539 D
1852 1458 D
2165 1377 D
2478 1317 D
2791 1319 D
3104 1246 D
3417 1208 D
stroke
grestore
end
showpage
}
\put(3054,1746){\makebox(0,0)[r]{Monte Carlo}}
\put(3054,1846){\makebox(0,0)[r]{variational, wavelets}}
\put(2008,2109){\makebox(0,0){Wavelet coefficients, anharmonic perturbation}}
\put(2008,51){\makebox(0,0){$\beta$}}
\put(100,1130){%
\special{ps: gsave currentpoint currentpoint translate
270 rotate neg exch neg exch translate}%
\makebox(0,0)[b]{\shortstack{${\cal A}^{(n)}$}}%
\special{ps: currentpoint grestore moveto}%
}
\put(3417,151){\makebox(0,0){1}}
\put(3104,151){\makebox(0,0){0.9}}
\put(2791,151){\makebox(0,0){0.8}}
\put(2478,151){\makebox(0,0){0.7}}
\put(2165,151){\makebox(0,0){0.6}}
\put(1852,151){\makebox(0,0){0.5}}
\put(1539,151){\makebox(0,0){0.4}}
\put(1226,151){\makebox(0,0){0.3}}
\put(913,151){\makebox(0,0){0.2}}
\put(600,151){\makebox(0,0){0.1}}
\put(540,1348){\makebox(0,0)[r]{1}}
\put(540,251){\makebox(0,0)[r]{0.1}}
\end{picture}
\caption{Wavelet coefficient fluctuations in the Gaussian model with an
anharmonicity ($m^2=1$, $g=0.2$).}
\label{fig8}
\end{figure}
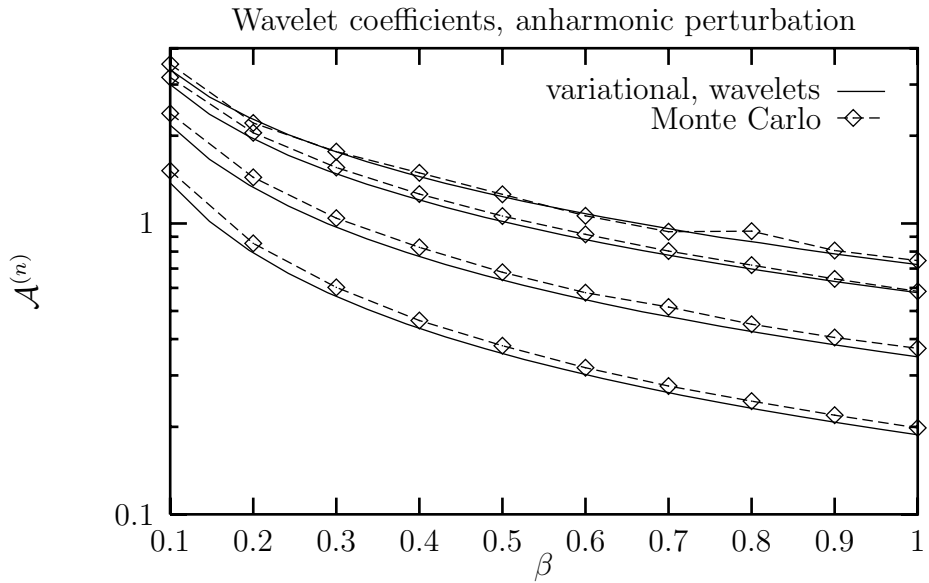

Small values of $g>0$ represent anharmonicities that manifest themselves in
the departures from the free-field result, comparable to perturbation
theory. Fig.~\ref{fig8} shows the wavelet coefficient fluctuations
calculated variationally and compared to the result of a Monte Carlo
calculation. Even where the departure from the free-field result is large
(at small $\beta$), the results are close.

\subsubsection{Landau-Ginzburg phase transition}

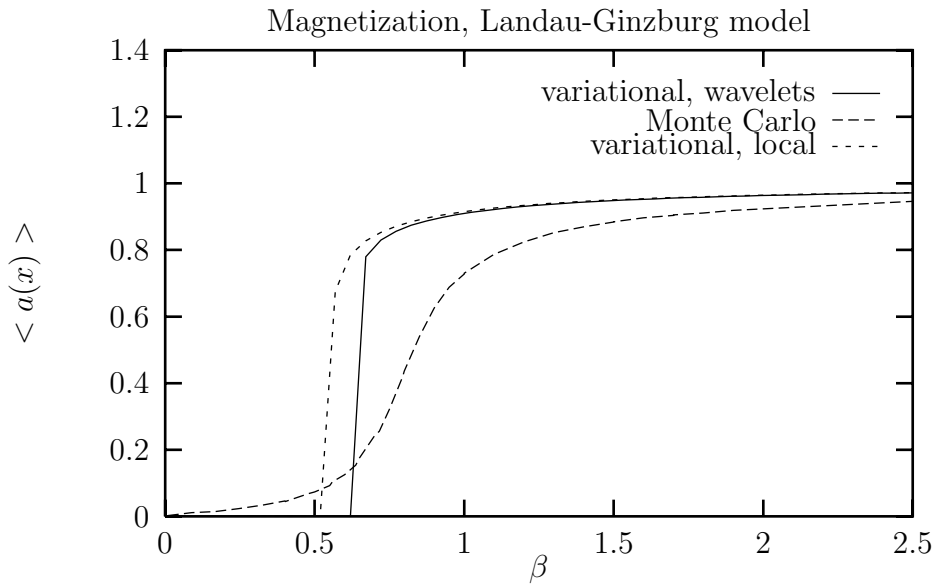
\begin{figure}
\setlength{\unitlength}{0.1bp}
\special{!
/gnudict 40 dict def
gnudict begin
/Color false def
/Solid false def
/gnulinewidth 5.000 def
/vshift -33 def
/dl {10 mul} def
/hpt 31.5 def
/vpt 31.5 def
/M {moveto} bind def
/L {lineto} bind def
/R {rmoveto} bind def
/V {rlineto} bind def
/vpt2 vpt 2 mul def
/hpt2 hpt 2 mul def
/Lshow { currentpoint stroke M
  0 vshift R show } def
/Rshow { currentpoint stroke M
  dup stringwidth pop neg vshift R show } def
/Cshow { currentpoint stroke M
  dup stringwidth pop -2 div vshift R show } def
/DL { Color {setrgbcolor Solid {pop []} if 0 setdash }
 {pop pop pop Solid {pop []} if 0 setdash} ifelse } def
/BL { stroke gnulinewidth 2 mul setlinewidth } def
/AL { stroke gnulinewidth 2 div setlinewidth } def
/PL { stroke gnulinewidth setlinewidth } def
/LTb { BL [] 0 0 0 DL } def
/LTa { AL [1 dl 2 dl] 0 setdash 0 0 0 setrgbcolor } def
/LT0 { PL [] 0 1 0 DL } def
/LT1 { PL [4 dl 2 dl] 0 0 1 DL } def
/LT2 { PL [2 dl 3 dl] 1 0 0 DL } def
/LT3 { PL [1 dl 1.5 dl] 1 0 1 DL } def
/LT4 { PL [5 dl 2 dl 1 dl 2 dl] 0 1 1 DL } def
/LT5 { PL [4 dl 3 dl 1 dl 3 dl] 1 1 0 DL } def
/LT6 { PL [2 dl 2 dl 2 dl 4 dl] 0 0 0 DL } def
/LT7 { PL [2 dl 2 dl 2 dl 2 dl 2 dl 4 dl] 1 0.3 0 DL } def
/LT8 { PL [2 dl 2 dl 2 dl 2 dl 2 dl 2 dl 2 dl 4 dl] 0.5 0.5 0.5 DL } def
/P { stroke [] 0 setdash
  currentlinewidth 2 div sub M
  0 currentlinewidth V stroke } def
/D { stroke [] 0 setdash 2 copy vpt add M
  hpt neg vpt neg V hpt vpt neg V
  hpt vpt V hpt neg vpt V closepath stroke
  P } def
/A { stroke [] 0 setdash vpt sub M 0 vpt2 V
  currentpoint stroke M
  hpt neg vpt neg R hpt2 0 V stroke
  } def
/B { stroke [] 0 setdash 2 copy exch hpt sub exch vpt add M
  0 vpt2 neg V hpt2 0 V 0 vpt2 V
  hpt2 neg 0 V closepath stroke
  P } def
/C { stroke [] 0 setdash exch hpt sub exch vpt add M
  hpt2 vpt2 neg V currentpoint stroke M
  hpt2 neg 0 R hpt2 vpt2 V stroke } def
/T { stroke [] 0 setdash 2 copy vpt 1.12 mul add M
  hpt neg vpt -1.62 mul V
  hpt 2 mul 0 V
  hpt neg vpt 1.62 mul V closepath stroke
  P  } def
/S { 2 copy A C} def
end
}
\begin{picture}(3600,2160)(0,0)
\special{"
gnudict begin
gsave
50 50 translate
0.100 0.100 scale
0 setgray
/Helvetica findfont 100 scalefont setfont
newpath
-500.000000 -500.000000 translate
LTa
600 251 M
2817 0 V
600 251 M
0 1758 V
LTb
600 251 M
63 0 V
2754 0 R
-63 0 V
600 502 M
63 0 V
2754 0 R
-63 0 V
600 753 M
63 0 V
2754 0 R
-63 0 V
600 1004 M
63 0 V
2754 0 R
-63 0 V
600 1256 M
63 0 V
2754 0 R
-63 0 V
600 1507 M
63 0 V
2754 0 R
-63 0 V
600 1758 M
63 0 V
2754 0 R
-63 0 V
600 2009 M
63 0 V
2754 0 R
-63 0 V
600 251 M
0 63 V
0 1695 R
0 -63 V
1163 251 M
0 63 V
0 1695 R
0 -63 V
1727 251 M
0 63 V
0 1695 R
0 -63 V
2290 251 M
0 63 V
0 1695 R
0 -63 V
2854 251 M
0 63 V
0 1695 R
0 -63 V
3417 251 M
0 63 V
0 1695 R
0 -63 V
600 251 M
2817 0 V
0 1758 V
-2817 0 V
600 251 L
LT0
3114 1846 M
180 0 V
611 251 M
58 0 V
57 0 V
57 0 V
57 0 V
58 0 V
57 0 V
57 0 V
57 0 V
58 0 V
57 0 V
57 0 V
57 0 V
58 979 V
57 63 V
57 33 V
57 23 V
58 16 V
57 13 V
57 11 V
57 9 V
58 7 V
57 7 V
57 6 V
58 5 V
57 4 V
57 4 V
57 4 V
58 3 V
57 3 V
57 3 V
57 2 V
58 2 V
57 3 V
57 2 V
57 1 V
58 2 V
57 2 V
57 1 V
57 2 V
58 1 V
57 1 V
57 1 V
57 2 V
58 1 V
57 1 V
57 1 V
57 1 V
58 0 V
57 1 V
LT1
3114 1746 M
180 0 V
611 254 M
89 11 V
88 4 V
88 11 V
89 13 V
86 16 V
2 -2 V
88 30 V
22 6 V
57 24 V
10 14 V
46 27 V
42 36 V
14 27 V
57 84 V
18 17 V
38 83 V
50 123 V
6 21 V
57 130 V
56 108 V
56 78 V
57 50 V
0 3 V
112 71 V
113 47 V
113 35 V
113 22 V
112 18 V
0 1 V
113 15 V
113 9 V
0 2 V
112 7 V
113 10 V
676 34 V
LT2
3114 1646 M
180 0 V
611 251 M
58 0 V
57 0 V
57 0 V
57 0 V
58 0 V
57 0 V
57 0 V
57 0 V
58 0 V
57 0 V
57 849 V
57 137 V
58 53 V
57 32 V
57 23 V
57 17 V
58 14 V
57 11 V
57 9 V
57 8 V
58 7 V
57 6 V
57 5 V
58 4 V
57 5 V
57 3 V
57 4 V
58 3 V
57 3 V
57 2 V
57 2 V
58 3 V
57 2 V
57 1 V
57 2 V
58 2 V
57 1 V
57 2 V
57 1 V
58 1 V
57 2 V
57 1 V
57 1 V
58 1 V
57 1 V
57 1 V
57 1 V
58 0 V
57 1 V
stroke
grestore
end
showpage
}
\put(3054,1646){\makebox(0,0)[r]{variational, local}}
\put(3054,1746){\makebox(0,0)[r]{Monte Carlo}}
\put(3054,1846){\makebox(0,0)[r]{variational, wavelets}}
\put(2008,2109){\makebox(0,0){Magnetization, Landau-Ginzburg model}}
\put(2008,51){\makebox(0,0){$\beta$}}
\put(100,1130){%
\special{ps: gsave currentpoint currentpoint translate
270 rotate neg exch neg exch translate}%
\makebox(0,0)[b]{\shortstack{$<a(x)>$}}%
\special{ps: currentpoint grestore moveto}%
}
\put(3417,151){\makebox(0,0){2.5}}
\put(2854,151){\makebox(0,0){2}}
\put(2290,151){\makebox(0,0){1.5}}
\put(1727,151){\makebox(0,0){1}}
\put(1163,151){\makebox(0,0){0.5}}
\put(600,151){\makebox(0,0){0}}
\put(540,2009){\makebox(0,0)[r]{1.4}}
\put(540,1758){\makebox(0,0)[r]{1.2}}
\put(540,1507){\makebox(0,0)[r]{1}}
\put(540,1256){\makebox(0,0)[r]{0.8}}
\put(540,1004){\makebox(0,0)[r]{0.6}}
\put(540,753){\makebox(0,0)[r]{0.4}}
\put(540,502){\makebox(0,0)[r]{0.2}}
\put(540,251){\makebox(0,0)[r]{0}}
\end{picture}
\caption{Magnetization in the Landau-Ginzburg model on a $32\times32$
lattice with $m^2=-4$, $g=2$, calculated both in a Monte Carlo simulation
and variationally using either wavelets and completely local fluctuations.}
\label{fig11}
\end{figure}

For $g>0$, $m<0$, the potential has the ``Mexican hat'' potential leading to
spontaneous symmetry breaking in mean field theory. Fluctuations change the
temperature of the symmetry-breaking phase transition from the mean-field
value $\beta=0$ to some finite value. In Fig.~\ref{fig11}, we have
calculated the magnetization variationally both using the wavelet basis and
the localized ansatz (\ref{posansatz}) and by a Monte Carlo calculation on
a $32\times32$ lattice with parameters $m^2=-4$ and $g=2$. In the Monte
Carlo calculation, an external field $h=0.05$ was applied. The finite
lattice size makes the phase transition less pronounced, but the inclusion
of nonlocal fluctuations in the wavelet ansatz lowers, as expected, the
phase transition temperature towards the result of the Monte Carlo
calculation.

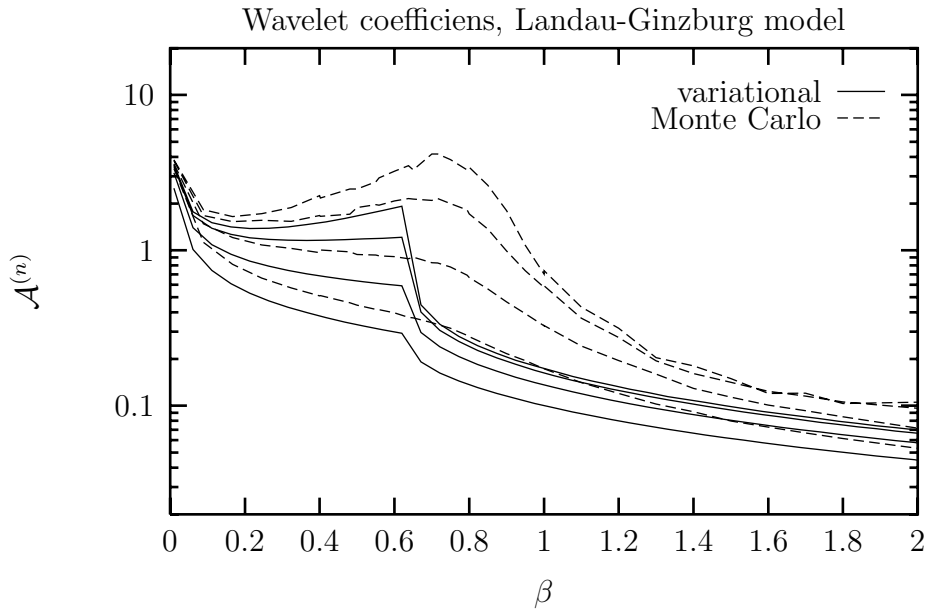
\begin{figure}
\setlength{\unitlength}{0.1bp}
\special{!
/gnudict 40 dict def
gnudict begin
/Color false def
/Solid false def
/gnulinewidth 5.000 def
/vshift -33 def
/dl {10 mul} def
/hpt 31.5 def
/vpt 31.5 def
/M {moveto} bind def
/L {lineto} bind def
/R {rmoveto} bind def
/V {rlineto} bind def
/vpt2 vpt 2 mul def
/hpt2 hpt 2 mul def
/Lshow { currentpoint stroke M
  0 vshift R show } def
/Rshow { currentpoint stroke M
  dup stringwidth pop neg vshift R show } def
/Cshow { currentpoint stroke M
  dup stringwidth pop -2 div vshift R show } def
/DL { Color {setrgbcolor Solid {pop []} if 0 setdash }
 {pop pop pop Solid {pop []} if 0 setdash} ifelse } def
/BL { stroke gnulinewidth 2 mul setlinewidth } def
/AL { stroke gnulinewidth 2 div setlinewidth } def
/PL { stroke gnulinewidth setlinewidth } def
/LTb { BL [] 0 0 0 DL } def
/LTa { AL [1 dl 2 dl] 0 setdash 0 0 0 setrgbcolor } def
/LT0 { PL [] 0 1 0 DL } def
/LT1 { PL [4 dl 2 dl] 0 0 1 DL } def
/LT2 { PL [2 dl 3 dl] 1 0 0 DL } def
/LT3 { PL [1 dl 1.5 dl] 1 0 1 DL } def
/LT4 { PL [5 dl 2 dl 1 dl 2 dl] 0 1 1 DL } def
/LT5 { PL [4 dl 3 dl 1 dl 3 dl] 1 1 0 DL } def
/LT6 { PL [2 dl 2 dl 2 dl 4 dl] 0 0 0 DL } def
/LT7 { PL [2 dl 2 dl 2 dl 2 dl 2 dl 4 dl] 1 0.3 0 DL } def
/LT8 { PL [2 dl 2 dl 2 dl 2 dl 2 dl 2 dl 2 dl 4 dl] 0.5 0.5 0.5 DL } def
/P { stroke [] 0 setdash
  currentlinewidth 2 div sub M
  0 currentlinewidth V stroke } def
/D { stroke [] 0 setdash 2 copy vpt add M
  hpt neg vpt neg V hpt vpt neg V
  hpt vpt V hpt neg vpt V closepath stroke
  P } def
/A { stroke [] 0 setdash vpt sub M 0 vpt2 V
  currentpoint stroke M
  hpt neg vpt neg R hpt2 0 V stroke
  } def
/B { stroke [] 0 setdash 2 copy exch hpt sub exch vpt add M
  0 vpt2 neg V hpt2 0 V 0 vpt2 V
  hpt2 neg 0 V closepath stroke
  P } def
/C { stroke [] 0 setdash exch hpt sub exch vpt add M
  hpt2 vpt2 neg V currentpoint stroke M
  hpt2 neg 0 R hpt2 vpt2 V stroke } def
/T { stroke [] 0 setdash 2 copy vpt 1.12 mul add M
  hpt neg vpt -1.62 mul V
  hpt 2 mul 0 V
  hpt neg vpt 1.62 mul V closepath stroke
  P  } def
/S { 2 copy A C} def
end
}
\begin{picture}(3600,2160)(0,0)
\special{"
gnudict begin
gsave
50 50 translate
0.100 0.100 scale
0 setgray
/Helvetica findfont 100 scalefont setfont
newpath
-500.000000 -500.000000 translate
LTa
600 251 M
0 1758 V
LTb
600 251 M
31 0 V
2786 0 R
-31 0 V
600 354 M
31 0 V
2786 0 R
-31 0 V
600 427 M
31 0 V
2786 0 R
-31 0 V
600 484 M
31 0 V
2786 0 R
-31 0 V
600 531 M
31 0 V
2786 0 R
-31 0 V
600 570 M
31 0 V
2786 0 R
-31 0 V
600 604 M
31 0 V
2786 0 R
-31 0 V
600 634 M
31 0 V
2786 0 R
-31 0 V
600 661 M
63 0 V
2754 0 R
-63 0 V
600 837 M
31 0 V
2786 0 R
-31 0 V
600 940 M
31 0 V
2786 0 R
-31 0 V
600 1013 M
31 0 V
2786 0 R
-31 0 V
600 1070 M
31 0 V
2786 0 R
-31 0 V
600 1117 M
31 0 V
2786 0 R
-31 0 V
600 1156 M
31 0 V
2786 0 R
-31 0 V
600 1190 M
31 0 V
2786 0 R
-31 0 V
600 1220 M
31 0 V
2786 0 R
-31 0 V
600 1247 M
63 0 V
2754 0 R
-63 0 V
600 1423 M
31 0 V
2786 0 R
-31 0 V
600 1526 M
31 0 V
2786 0 R
-31 0 V
600 1599 M
31 0 V
2786 0 R
-31 0 V
600 1656 M
31 0 V
2786 0 R
-31 0 V
600 1703 M
31 0 V
2786 0 R
-31 0 V
600 1742 M
31 0 V
2786 0 R
-31 0 V
600 1776 M
31 0 V
2786 0 R
-31 0 V
600 1806 M
31 0 V
2786 0 R
-31 0 V
600 1833 M
63 0 V
2754 0 R
-63 0 V
600 2009 M
31 0 V
2786 0 R
-31 0 V
600 251 M
0 63 V
0 1695 R
0 -63 V
882 251 M
0 63 V
0 1695 R
0 -63 V
1163 251 M
0 63 V
0 1695 R
0 -63 V
1445 251 M
0 63 V
0 1695 R
0 -63 V
1727 251 M
0 63 V
0 1695 R
0 -63 V
2009 251 M
0 63 V
0 1695 R
0 -63 V
2290 251 M
0 63 V
0 1695 R
0 -63 V
2572 251 M
0 63 V
0 1695 R
0 -63 V
2854 251 M
0 63 V
0 1695 R
0 -63 V
3135 251 M
0 63 V
0 1695 R
0 -63 V
3417 251 M
0 63 V
0 1695 R
0 -63 V
600 251 M
2817 0 V
0 1758 V
-2817 0 V
600 251 L
LT0
3114 1846 M
180 0 V
614 1481 M
72 -231 V
71 -79 V
72 -50 V
71 -36 V
72 -29 V
72 -24 V
71 -20 V
72 -19 V
71 -17 V
72 -15 V
71 -14 V
72 -13 V
72 -108 V
71 -42 V
72 -30 V
71 -25 V
72 -22 V
71 -19 V
72 -19 V
72 -16 V
71 -16 V
72 -14 V
71 -14 V
72 -13 V
71 -12 V
72 -12 V
72 -11 V
71 -11 V
72 -10 V
71 -9 V
72 -10 V
71 -9 V
72 -9 V
72 -8 V
71 -8 V
72 -8 V
71 -8 V
72 -7 V
72 -7 V
11 -2 V
614 1539 M
72 -207 V
71 -64 V
72 -36 V
71 -26 V
72 -19 V
72 -16 V
71 -13 V
72 -11 V
71 -10 V
72 -9 V
71 -8 V
72 -7 V
72 -176 V
71 -54 V
72 -37 V
71 -30 V
72 -25 V
71 -23 V
72 -20 V
72 -18 V
71 -17 V
72 -16 V
71 -15 V
72 -14 V
71 -13 V
72 -12 V
72 -12 V
71 -11 V
72 -11 V
71 -10 V
72 -10 V
71 -10 V
72 -9 V
72 -8 V
71 -9 V
72 -8 V
71 -8 V
72 -8 V
72 -7 V
11 -2 V
614 1563 M
72 -185 V
71 -48 V
72 -23 V
71 -13 V
72 -6 V
72 -3 V
71 -1 V
72 1 V
71 2 V
72 3 V
71 2 V
72 4 V
72 -282 V
71 -69 V
72 -42 V
71 -34 V
72 -28 V
71 -24 V
72 -23 V
72 -20 V
71 -18 V
72 -16 V
71 -16 V
72 -14 V
71 -14 V
72 -13 V
72 -12 V
71 -12 V
72 -11 V
71 -11 V
72 -10 V
71 -9 V
72 -10 V
72 -10 V
71 -8 V
72 -9 V
71 -7 V
72 -9 V
72 -7 V
11 -1 V
614 1569 M
72 -177 V
71 -41 V
72 -16 V
71 -6 V
72 1 V
72 5 V
71 9 V
72 10 V
71 13 V
72 15 V
71 15 V
72 16 V
72 -372 V
71 -73 V
72 -46 V
71 -35 V
72 -29 V
71 -26 V
72 -21 V
72 -20 V
71 -20 V
72 -17 V
71 -15 V
72 -16 V
71 -13 V
72 -15 V
72 -11 V
71 -11 V
72 -12 V
71 -11 V
72 -11 V
71 -9 V
72 -9 V
72 -9 V
71 -10 V
72 -8 V
71 -9 V
72 -7 V
72 -8 V
11 -1 V
LT1
3114 1746 M
180 0 V
614 1555 M
725 1277 L
110 -83 V
110 -52 V
111 -37 V
107 -30 V
3 1 V
111 -27 V
27 -9 V
71 -16 V
12 -1 V
58 -12 V
53 -17 V
18 -2 V
70 -18 V
22 -7 V
48 -14 V
63 -27 V
8 -5 V
70 -32 V
71 -32 V
70 -26 V
71 -31 V
140 -52 V
141 -41 V
141 -40 V
141 -29 V
141 -35 V
141 -23 V
140 -22 V
141 -21 V
282 -37 V
0 1 V
614 1573 M
725 1347 L
110 -52 V
110 -28 V
111 -13 V
107 -16 V
3 10 V
111 -7 V
27 -10 V
71 -2 V
12 -3 V
58 -3 V
53 -8 V
18 6 V
70 -21 V
22 -1 V
48 -15 V
63 -31 V
8 -9 V
70 -43 V
71 -49 V
70 -44 V
71 -46 V
0 1 V
140 -77 V
141 -55 V
141 -51 V
141 -54 V
141 -33 V
141 -30 V
140 -21 V
141 -23 V
282 -43 V
0 2 V
614 1585 M
725 1378 L
110 -23 V
110 5 V
111 -4 V
107 22 V
3 -3 V
111 8 V
27 25 V
71 11 V
12 -2 V
58 16 V
53 9 V
18 -2 V
70 -5 V
22 7 V
48 -19 V
63 -24 V
8 -15 V
70 -57 V
71 -83 V
70 -74 V
71 -63 V
0 6 V
2149 991 L
141 -73 V
141 -88 V
141 -48 V
141 -33 V
141 -34 V
140 -18 V
141 -22 V
282 -24 V
0 20 V
614 1588 M
725 1400 L
110 -26 V
110 11 V
111 23 V
107 45 V
3 -9 V
111 34 V
27 0 V
71 27 V
12 15 V
58 25 V
53 21 V
18 -16 V
70 60 V
22 0 V
48 -21 V
63 -39 V
8 10 V
70 -68 V
71 -96 V
70 -130 V
71 -109 V
0 13 V
140 -137 V
141 -80 V
2431 842 L
141 -31 V
141 -48 V
141 -56 V
140 1 V
141 -39 V
282 5 V
0 388 V
stroke
grestore
end
showpage
}
\put(3054,1746){\makebox(0,0)[r]{Monte Carlo}}
\put(3054,1846){\makebox(0,0)[r]{variational}}
\put(2008,2109){\makebox(0,0){Wavelet coefficiens, Landau-Ginzburg model}}
\put(2008,-49){\makebox(0,0){$\beta$}}
\put(100,1130){%
\special{ps: gsave currentpoint currentpoint translate
270 rotate neg exch neg exch translate}%
\makebox(0,0)[b]{\shortstack{${\cal A}^{(n)}$}}%
\special{ps: currentpoint grestore moveto}%
}
\put(3417,151){\makebox(0,0){2}}
\put(3135,151){\makebox(0,0){1.8}}
\put(2854,151){\makebox(0,0){1.6}}
\put(2572,151){\makebox(0,0){1.4}}
\put(2290,151){\makebox(0,0){1.2}}
\put(2009,151){\makebox(0,0){1}}
\put(1727,151){\makebox(0,0){0.8}}
\put(1445,151){\makebox(0,0){0.6}}
\put(1163,151){\makebox(0,0){0.4}}
\put(882,151){\makebox(0,0){0.2}}
\put(600,151){\makebox(0,0){0}}
\put(540,1833){\makebox(0,0)[r]{10}}
\put(540,1247){\makebox(0,0)[r]{1}}
\put(540,661){\makebox(0,0)[r]{0.1}}
\end{picture}
\caption{Wavelet coefficient fluctuations in the Landau-Ginzburg model,
calculated variationally and in a Monte Carlo simulation.}
\label{fig10}
\end{figure}

Fig.~\ref{fig10} shows the wavelet fluctuations calculated variationally and
in the Monte Carlo simulation. The variational wavelet ansatz gives a
faithful representation of the overall features. In particular, the wavelet
coefficients both in the Monte Carlo and the variational calculation develop
an approximate power scaling near the phase transition, indicating the
presence of fluctuations at all scales. In contrary, fluctuations are strong
but without scaling in the disordered phase where only short-range
correlations persist. In the low-temperature phase, the scaling is that of
the massive Gaussian model.

%

\section{Conclusions}

We have shown how wavelets can be used to describe fluctuations in a
statistical field theory. Using a rather small number of variational
parameters, an approximate description of fluctuations in the Gaussian, the
$XY$, and the Landau-Ginzburg model has been obtained. The wavelet
coefficients compactly express the scaling features of the field theory in
only a few numbers. Comparison to Monte Carlo results leads us to expect
that wavelets are a sufficiently efficient basis to merit further
investigation of their use in renormalization group transformations.

As our results are only semiquantitative, we have refrained from calculating
critical exponents directly. Since the results are arrived at numerically,
it is very difficult to extract critical exponents by differentiating
thermodynamic quantities. Once an implementation of the renormalization
group in wavelet space is in place, they can be extracted much easier.

Such renormalization group transformations can be implemented by
considering the wavelet coefficients as the basic dynamical variables of
the system and then integrating them out scale by scale. The wavelet
transform has the advantage of having a hierarchy built-in, unlike lattices
where the hierarchy is imposed {\em ad hoc}. The renormalization group
transformation can be made either in (wavelet) perturbation theory, using a
variational principle, or directly when restricting the model to a
hierarchical model \cite{Savv}. Work in these directions is in progress.

We would like to thank T.~Bir\'o, P.~Carruthers, M.~Greiner, and G.~Savvidy
for fruitful discussions during the progress of this project, and the
German National Scholarship Foundation for its support.

\end{document}